\documentclass{siamltex}
\usepackage{makeidx}
\usepackage{amsfonts}
\usepackage{amsmath}
\usepackage{amssymb}
\usepackage{amscd}
\usepackage[dvips]{graphicx}
\usepackage[noend]{algorithmic}
\usepackage{algorithm}
\usepackage{comment}
\usepackage{color}
\usepackage{array,amssymb,amsmath,amsbsy}
\usepackage{enumerate}
\usepackage[bottom]{footmisc}
\usepackage[all,cmtip]{xy}
\usepackage{hyperref}
\usepackage{xspace}
\usepackage{wrapfig}
\usepackage{simplemargins}
\setallmargins{1in}
\def\Aut{{\rm Aut}}

\def\O{\calO}

\def\bo{\partial} 
\def\int{\mathring} 

\def\listiso{
  \mathrel{\vbox{\offinterlineskip\ialign{%
     \hfil##\hfil\cr
     $\scriptstyle\frakL$\cr
     \noalign{\kern 0.1ex}
     $\longrightarrow$\cr
  }}}}
\def\coliso{
  \mathrel{\vbox{\offinterlineskip\ialign{%
     \hfil##\hfil\cr
     $\scriptstyle c$\cr
     \noalign{\kern 0.1ex}
     $\longrightarrow$\cr
  }}}}

\def\rcover{\textsc{RegularCover}\xspace}
\def\cover{\textsc{$H$-Cover}\xspace}
\def\hom{\textsc{$H$-Hom}\xspace}
\def\autprob{\textsc{AutGroup}\xspace}
\def\ivmatch{{\sffamily IV}-\textsc{Matching}\xspace}
\def\ivsubgraph{{\sffamily IV}-subgraph\xspace}

\def\lifting{\textsc{RegularLifting}\xspace}
\def\quotient{\textsc{RegularQuotient}\xspace}
\def\foldbracketrcover#1{\textsc{$#1$-Fold(Regular)Cover}\xspace}
\def\foldrcover#1{\textsc{$#1$-FoldRegularCover}\xspace}
\def\foldcover#1{\textsc{$#1$-FoldCover}\xspace}

\def\gi{\textsc{GraphIso}\xspace}
\def\lgi{\textsc{ListIso}\xspace}

\def\cgi{\textsc{ColorIso}\xspace}

\newenvironment{packed_enum}{
	\begin{enumerate}
		\setlength{\itemsep}{1pt}
	    \setlength{\parskip}{0pt}
		\setlength{\parsep}{0pt}
}{\end{enumerate}}

\newenvironment{packed_itemize}{
	\begin{itemize}
		\setlength{\itemsep}{1pt}
	    \setlength{\parskip}{0pt}
		\setlength{\parsep}{0pt}
}{\end{itemize}}

\newenvironment{packed_head_enum}[1]{
	\begin{enumerate}[#1]
		\setlength{\itemsep}{1pt}
	    \setlength{\parskip}{0pt}
		\setlength{\parsep}{0pt}
}{\end{enumerate}}

\newtheorem{problem}[theorem]{Problem}

\newcommand{\heading}[1]{\medskip\par\noindent{\bf #1}}

\def\computationproblem#1#2#3#4{
	\vskip 1ex
	\begin{center}
	\fbox{\begin{tabular}{rp{#4}}
	{\bf Problem:}&#1\\
	{\bf Input:}&#2\\
	{\bf Output:}&#3\\
	\end{tabular}}
	\end{center}
	\vskip 1ex
}

\def\bV{\boldsymbol{V}}
\def\bE{\boldsymbol{E}}
\def\bH{\boldsymbol{H}}
\def\bP{\boldsymbol{P}}
\def\bv{\boldsymbol{v}}
\def\be{\boldsymbol{e}}
\def\bh{\boldsymbol{h}}
\def\bhatv{\boldsymbol{\hat v}}
\def\bhate{\boldsymbol{\hat e}}

\def\gS{\mathbb{S}} \def\gC{\mathbb{C}} \def\gA{\mathbb{A}} \def\gD{\mathbb{D}}

\def\calA{{\cal A}}  \def\calC{{\cal C}} 
   \def\calH{{\cal H}}
   
\def\calM{{\cal M}}  \def\calO{{\cal O}} \def\calP{{\cal P}}
 \def\calR{{\cal R}}

   \def\frakL{{\mathfrak L}}

\def\cNP{\hbox{\rm \sffamily NP}\xspace}
\def\cGI{\hbox{\rm \sffamily GI}\xspace}

\def\cFPT{\hbox{\rm \sffamily FPT}\xspace}
\def\ccoAM{\hbox{\rm \sffamily coAM}\xspace}

\title{Algorithmic Aspects of Regular Graph Covers\thanks{
This paper continues the research started in ICALP 2014~\cite{fkkn} and extends its results. For a
structural diagram visualizing our results, see
\url{http://pavel.klavik.cz/orgpad/regular_covers.html} (supported for Firefox and Google Chrome).
This work was initiated during workshops Algebraic, Topological and Complexity Aspects of Graph
Covers (ATCACG). The authors are supported by CE-ITI (P202/12/G061 of GA\v{C}R).  The first author
is also supported by the project Kontakt LH12095, the second and the third authors by Charles
University as GAUK 196213, the fourth author by the Ministry of Education of the Slovak Republic,
the grant VEGA 1/0150/14, by Project LO1506 of the Czech Ministry of Education and by the project
APVV-15-0220.}}
\author{Ji\v{r}\'i Fiala\footnotemark[2]
		\and Pavel Klav\'ik\footnotemark[3]
		\and Jan Kratochv\'il\footnotemark[2]
		and Roman Nedela\footnotemark[4]\ $^{,}$\footnotemark[5]
		}
\begin{document}
\maketitle

\renewcommand{\thefootnote}{\fnsymbol{footnote}}
\footnotetext[2]{Department of Applied Mathematics, Faculty of Mathematics and Physics, Charles
University, Malostransk{\'e} n{\'a}m{\v e}st{\'\i} 25, 118 00 Prague, Czech Republic.\\
E-mails: \texttt{\{fiala,honza\}@kam.mff.cuni.cz}.}
\footnotetext[3]{Computer Science Institute, Faculty of Mathematics and Physics, Charles University,
Malostransk{\'e} n{\'a}m{\v e}st{\'\i} 25, 118 00 Prague, Czech Republic.  E-mail:
\texttt{klavik@iuuk.mff.cuni.cz}.}
\footnotetext[4]{Institute of Mathematics and Computer Science SAS,
\v{D}umbierska 1, 974 11 Bansk\'a Bystrica, Slovak republic. Email: \texttt{nedela@savbb.sk}.}
\footnotetext[5]{European Centre of Excellence NTIS, University of West Bohemia, Pilsen, Czech Republic.}
\renewcommand{\thefootnote}{\arabic{footnote}}

\begin{abstract}
A graph $G$ \emph{covers} a graph $H$ if there exists a locally bijective homomorphism from $G$ to
$H$. We deal with \emph{regular covers} where this homomorphism is prescribed by the action of a
semiregular subgroup of $\Aut(G)$.
We study \emph{computational aspects} of regular covers that have not been addressed
before. The decision problem $\rcover$ asks for given graphs $G$ and $H$ whether $G$ regularly
covers $H$.  When $|H|=1$, this problem becomes Cayley graph recognition for which the complexity is
still unresolved. Another special case arises for $|G| = |H|$ when it becomes the graph isomorphism
problem. 

Our main result is an involved \cFPT algorithm solving $\rcover$ for planar inputs $G$ in time
$\O^*(2^{\be(H)/2})$ where $\be(H)$ denotes the number of edges of $H$. The algorithm is based on
dynamic programming and employs theoretical results proved in a related structural paper. Further,
when $G$ is 3-connected, $H$ is 2-connected or the ratio $|G|/|H|$ is an odd integer, we can solve
the problem $\rcover$ in polynomial time. In comparison, B\'ilka et al. (2011) proved that testing
general graph covers is \cNP-complete for planar inputs $G$ when $H$ is a small fixed graph such as
$K_4$ or $K_5$. 
\end{abstract}

\begin{keywords}
regular graph covers, planar graphs, FPT algorithm, computational complexity, graph isomorphism
problem, Cayley graph recognition
\end{keywords}


\section{Introduction} \label{sec:introduction}

The notion of \emph{covering} originates in topology as a notion of local similarity of two
topological spaces. For instance, consider the unit circle and the real line. Globally, these two
spaces are not the same, they have different properties, different fundamental groups, etc.
But when we restrict ourselves to a small part of the circle, it looks the same as a small part
of the real line; more precisely the two spaces are locally homeomorphic, and thus they share the
local properties. The notion of covering formalizes this property of two spaces being
\emph{locally the same}.

\begin{figure}[t!]
\centering
\includegraphics{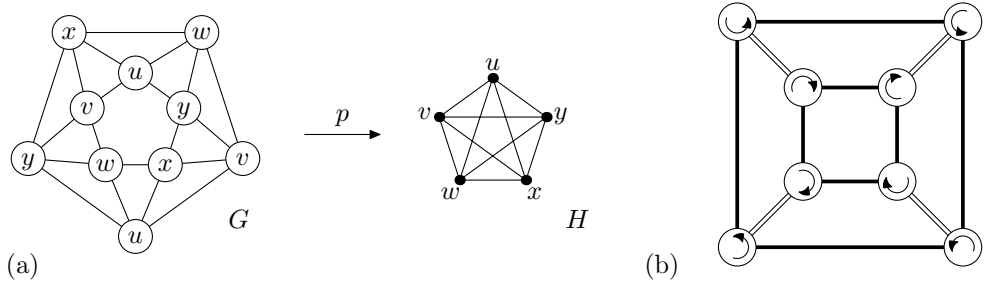}
\caption{(a) A covering projection $p$ from a graph $G$ to a graph $H$. (b) The Cayley
graph of the dihedral group $\gD_4$ generated by the $90^\circ$ rotations (in black) and the
reflection around the $x$-axis (in white).}
\label{fig:big_picture}
\end{figure}

Suppose that we have two topological spaces: a big one $G$ and a small one $H$. We say that $G$
\emph{covers} $H$ if there exists an epimorphism called a \emph{covering projection} $p : G \to H$
which locally preserves the structure of $G$.  For instance, the mapping $p(x) = (\cos x,\sin x)$
from the real line to the unit circle is a covering projection. The existence of a covering
projection ensures that $G$ looks locally the same as $H$; see Fig.~\ref{fig:big_picture}a.

In this paper, we study coverings of graphs in a more restricting version called \emph{regular
covering}, for which the covering projection is described by an action of a group; see
Section~\ref{sec:preliminaries} for the formal definition. If $G$ regularly covers $H$, then we say
that $H$ is a \emph{(regular) quotient} of $G$.

Negami's Theorem~\cite{negami}, stating that all regular quotients of planar graphs can be embedded
into the projective plane, is one of the oldest results in topological graph theory. Therefore, we
have decided to initiate the study of computational complexity of regular graph covers with planar
graphs.

\subsection{Applications of Graph Coverings}

Suppose that $G$ covers $H$ and we have some information about one of the objects. How much
knowledge does translate to the other object? It turns out that quite a lot, and this makes
covering a powerful technique with many diverse applications. The big advantage of regular coverings
is that they can be efficiently described and many properties easily translate between the objects.
We sketch some applications now.

\heading{Powerful Constructions.} The reverse of covering called \emph{lifting} can be applied to
small objects in order to construct large objects of desired properties. For instance, the
well-known Cayley graphs are large objects which can be described easily by a few elements of a
group. Let $G$ be a Cayley graph generated by elements $g_1,\dots,g_e$ of a group $\Gamma$. The
vertices of $G$ correspond to the elements of $\Gamma$ and the edges are described by actions of
$g_1,\dots,g_e$ on $\Gamma$ by left multiplication; each $g_i$ defines a permutation on $\Gamma$ and
we put edges along the cycles of this permutation. See Fig.~\ref{fig:big_picture}b for an example.
Cayley graphs were originally invented to study the structure of groups~\cite{cayley}.

\begin{figure}[b!]
\centering
\includegraphics{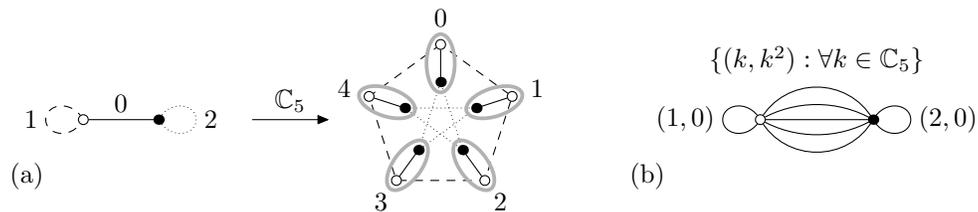}
\caption{(a) A construction of the Petersen graph by lifting with the group $\gC_5$. (b) By lifting
the described graph with the group $\gC_5^2$, we get the Hoffman-Singleton graph. The five parallel
edges are labeled $(0,0)$, $(1,1)$, $(2,4)$, $(3,4)$ and $(4,1)$.}
\label{fig:lifting_graphs}
\end{figure}

In the language of coverings, every Cayley graph $G$ can be described as a lift of a one vertex
graph $H$ with $e$ loops and half-edges attached labeled $g_1,\dots,g_e$. Regular covers can be
viewed as a generalization of Cayley graphs where the small graph $H$ can contain more than one
vertex. For example, the famous Petersen graph can be constructed as a lift of a two-vertex graph
$H$ in Fig.~\ref{fig:lifting_graphs}a. These two vertices are necessary as it is known that Petersen
graph is not a Cayley graph. Figure~\ref{fig:lifting_graphs}b shows a simple
construction~\cite{hoffman_construction,siagiova} of the Hoffman-Singleton
graph~\cite{hoffman_singleton} which is a 7-regular graph with 50 vertices. 

The Petersen and the Hoffman-Singleton graphs are extremal graphs for the \emph{degree-diameter}
problem: given integers $d$ and $k$, find a maximal graph $G$ with diameter $d$ and degree $k$. In
general, the size of $G$ is not known. Many currently best constructions were obtained using the
covering techniques~\cite{miller2005moore}. 

Further applications employ the fact that nowhere-zero flows, vertex and edge colorings, eigenvalues
and other graph invariants lift along a covering projection. Two main applications of constructions
of lifts are the solution of the Heawood map coloring problem~\cite{ringel_youngs,gross_tucker} and
constructions of arbitrarily large highly symmetrical graphs~\cite{biggs}.

\heading{Models of Local Computation.} These and similar constructions have many practical
applications in designing highly efficient computer networks~\cite{network0,network1,network2,%
network3,network4,network5,network6,network7}, since these networks can be efficiently
described/constructed and have many strong properties. In particular, networks based on covers of
simple graphs allow fast parallelization of computation as described e.g.
in~\cite{bodlaender,Ang1,Ang2}.

\heading{Simplifying Objects.} Regular coverings can be also applied in the opposite way, to project
big objects onto smaller ones while preserving some properties. One way is to represent a class of
objects satisfying some properties as quotients of the universal object of this property.  For
instance, this was used in the study of arc-transitive cubic graphs~\cite{goldschmidt}, and the key
point is that universal objects are much easier to work with. This idea is commonly used in fields
such as the theory of Riemann surfaces~\cite{farkas_kra} and theoretical physics~\cite{katanaev}.

\subsection{Regular Covering Testing}

Despite all described applications, the computational complexity of regular covering was not yet
studied. In this paper, we initiate the study of the following computational problem.

\computationproblem
{\rcover}
{Connected graphs $G$ and $H$.}
{Does $G$ regularly cover $H$?}
{4.25cm}

For a fixed graph $H$, the computational complexity of \rcover was first asked as an open problem by
Abello et al.~\cite{AFS}: ``Are there graphs $H$ for which the problem of determining if an input
graph $G$ is a regular cover of $H$ is \cNP-hard?'' Currently, no \cNP-hardness reduction is known
for \rcover, even when $H$ is a part of the input.  Our main result shows that if $G$ is planar, no
such graph $H$ exists.  We use the complexity notation $f = \O^*(g)$ which omits polynomial
factors. We establish the following \cFPT algorithm:

\begin{theorem} \label{thm:planar_rcover}
For planar graphs $G$, the \rcover\ problem can be solved in time $\O^*(2^{\be(H)/2})$, where $\be(H)$
is the number of edges of $H$.
\end{theorem}

\subsection{Related Computational Problems}

We discuss other computational problems related to \rcover.  The notion of regular covers builds a
bridge between two seemingly different problems: Cayley graph recognition and the graph isomorphism
problem. 

\heading{Covering Testing.} The complexity of general covering was widely studied before, pioneered
by Bodlaender~\cite{bodlaender} in the context of networks of processors in parallel computing.
Abello et al.~\cite{AFS} introduced the \cover problem which asks for an input graph $G$ whether it
covers a fixed graph $H$. Unless $H$ is very simple, the problem turned out to be mostly
\cNP-complete, the general complexity is still unresolved but the papers~\cite{kratochvil97,fiala00}
show that it is \cNP-complete for every $r$-regular graph $H$ where $r \ge 3$. For a survey of the
complexity results, see~\cite{FK}.

We try to understand how much the additional algebraic structure of regular covering changes the
computational complexity.  For planar inputs $G$, the change is significant: the problem \cover
remains \cNP-complete for several small fixed graphs $H$ (such as $K_4$,
$K_5$)~\cite{planar_covers}, while \rcover can be solved in polynomial time for every fixed graph
$H$ by Theorem~\ref{thm:planar_rcover}.

\heading{Cayley Graphs Testing.}
If the graph $H$ consists of a single vertex with attached loops and half-edges, it corresponds to
Cayley graph recognition whose computational complexity is widely open. No hardness results are
known and a polynomial-time algorithm is known only for recognition of circulant
graphs~\cite{circulant_recognition}. In contrast, if $H$ consists of a vertex with three half-edges
attached, then $G$ covers $H$ if and only if $G$ is a cubic 3-edge-colorable graph, so \cover is
\cNP-complete~\cite{holyer}.

The reader may notice that Theorem~\ref{thm:planar_rcover} gives a polynomial time algorithm to
recognize planar Cayley graphs. The input is a $k$-regular planar graph $G$, for $k \le 5$. We test
\rcover for all graphs $H$ which a single vertex of degree $k$. Unfortunately, finite planar Cayley
graphs $G$ are very limited: either $G$ is a cycle, or $G$ is 3-connected. Therefore, $\Aut(G)$ is a
\emph{spherical group} which is very simple. Therefore, $G$ is either finite (with $v(G) \le 120$),
representing one of the sporadic groups (for instance, a truncated dodecahedron is a Cayley graph of
$\gA_5$), or very simple (a cycle, a prism, an antiprism, e.g.).

\heading{Graph Isomorphism Problem.}
The other extreme is when both graphs $G$ and $H$ have the same size, for which \rcover is the
famous graph isomorphism problem (\gi). Graph isomorphism belongs to \cNP, it is unlikely
\cNP-complete, however no polynomial-time algorithm is known and Babai~\cite{babai_quasipoly}
recently proved that it can be solved in quasipolynomial time. Also, polynomial-time algorithms for
\gi are known for many graph classes and parameters; see~\cite{kkz} for an overview. Since \rcover
generalizes \gi, we cannot hope to solve it in polynomial time (unless solving \gi as well). It is
natural to ask which results and techniques for \gi translate to \rcover. Our results show that some
technique for planar graphs translate, but the \rcover problem is significantly more involved.

Theoretical motivation for studying the graph isomorphism problem is very similar to \rcover. For
practical instances, one can solve \gi very efficiently using various heuristics. But
polynomial-time algorithm working for all graphs is not known and it is very desirable to understand
the complexity of \gi. It is known that testing graph isomorphism is equivalent to testing
isomorphism of general mathematical structures~\cite{hedrlin}.  The notion of isomorphism is widely
used in mathematics when one wants to show that two seemingly different structures are the same. One
proceeds by guessing a mapping and proving that this mapping is an isomorphism. The natural
complexity question is whether there is a better algorithmic way to derive an isomorphism.
Similarly, regular covering is a well-known mathematical notion which is algorithmically interesting
and not understood.

\heading{Computing Automorphism Groups.}
A regular covering is described by a semiregular subgroup of the automorphism group
$\Aut(G)$. We denote the computational problem of finding generators of $\Aut(G)$ by
\autprob. Since a good understanding of $\Aut(G)$ is needed to solve \rcover, it is closely related
to \autprob.

It is known that \gi can be reduced to \autprob (which forms the foundation of group theory
techniques used to attack the graph isomorphism problem,
e.g.,~\cite{luks1982isomorphism,babai_quasipoly}), moreover \autprob can be solved by $\O(n^3)$
instances of \gi~\cite{mathon_isocount}. Surprisingly not much is known about automorphism groups of
restricted classes of graphs. Jordan~\cite{jordan} gave an inductive characterization of automorphism
groups of trees as the class of groups closed under direct product and wreath product with symmetric
groups. Babai~\cite{babai1975automorphism,babai1996automorphism} described automorphism groups of
planar graphs. Recently, Jordan-like characterizations of automorphism groups of interval, circle
and permutation graphs are given in~\cite{kz,kz15}. The \autprob problem can be solved in linear
time for trees and interval graphs~\cite{colbourn_booth}, in linear time for permutation
graphs~\cite{kz15}, and in polynomial time for circle graphs~\cite{kz15}.

Our description of semiregular actions on planar graphs in~\cite{fkkn16} was generalized
in~\cite{knz} to describe a Jordan-like characterization of automorphism groups of planar graphs,
which is much more detailed than Babai's description in~\cite{babai1975automorphism}. It also implies a
quadratic-time algorithm for \autprob of planar graphs (which likely can be improved to linear
time), faster than the best previous trivial $\O(n^4)$ algorithm by
combining~\cite{hopcroft1974linear,mathon_isocount}.

\heading{List Restricted Isomorphism Problem.}
Let $G$ and $H$ be graphs and the vertices of $G$ be \emph{equipped by lists:} for each $u \in \bV(G)$, we
have $\frakL(u) \subseteq \bV(H)$.  An isomorphism $\pi : G \to H$ is called \emph{list-compatible} if
for every $u \in \bV(G)$, we have $\pi(u) \in \frakL(u)$. The existence of a list-compatible
isomorphism is denoted by $G \listiso H$.

\computationproblem
{\lgi}
{Graph $G$ and $H$, and for each $u \in \bV(G)$ a list $\frakL(u) \subseteq \bV(H)$.}
{Does $G \listiso H$?}
{9.5cm}

This problem was first introduced by Lubiw~\cite{lubiw} and proved to be \cNP-complete, even in the
following restricted setting.

\begin{theorem}[Lubiw~\cite{lubiw}] \label{thm:lubiw}
Testing existence of a fixed-point free involutory automorphism is \cNP-complete.
\end{theorem}

But only the above result of~\cite{lubiw} is cited while the \lgi problem was forgotten. We have
rediscovered \lgi since it was solved in a subroutine in our algorithm of
Theorem~\ref{thm:planar_rcover} for 3-connected planar and projectively planar graphs, for which it
can be solved in polynomial time using~\cite{pp_iso}; see~\cite{fkkn}. Our paper gives a nice
motivation for \lgi, leading Klav\'{\i}k et al.~\cite{kkz} to study it for many restricted
graph classes and parameters. In particular, \lgi can be solved in polynomial time for graphs of
bounded genus and bounded treewidth~\cite{kkz}.

We also consider special instances called \cgi in which both graphs $G$ and $H$ are colored and we
ask for existence of a color-preserving isomorphism, denoted $G \coliso H$. Unlike \lgi, the \cgi
problem is a well known problem which is polynomial-time equivalent to \gi.

\heading{Homomorphisms and CSP.}
Since regular covering is a locally bijective homomorphism, we give an overview of complexity
results concerning homomorphisms.  Hell and Ne\v{s}et\v{r}il~\cite{hell_nesetril} studied the
problem \hom which asks whether there exists a homomorphism between an input graph $G$ and a fixed
graph $H$. Their celebrated dichotomy result for simple graphs states that the problem \hom is
polynomially solvable if $H$ is bipartite, and it is \cNP-complete otherwise. Homomorphisms can be
described in the language of constraint satisfaction (CSP), and the famous dichotomy
conjecture~\cite{csp_dichotomy} claims that every CSP is either polynomially solvable, or
\cNP-complete.  

\subsection{Other Covering Problems}

We introduce and discuss several other problems related to (regular) graph covering.

\heading{Lifting and Quotients.}
In the \rcover problem, the input gives two graphs $G$ and $H$.  For the following problems, the
input specifies only one graph and we ask for existence of the other graph:

\computationproblem
{\lifting}
{A connected graph $H$ and an integer $k$.}
{Does there exists a graph $G$ regularly covering $H$ such that $|G| = k|H|$?}
{8cm}

\computationproblem
{\quotient}
{A connected graph $G$ and an integer $k$.}
{Does there exists a graph $H$ regularly covered by $G$ such that $|H| = {|G| \over k}$?}
{8cm}

Concerning \lifting, the answer is always positive. The theory of covering describes a technique
called voltage assignment which can be applied to generate all $k$-folds $G$.  We do not deal with
lifting in this paper, but there are nevertheless many interesting computational questions with
applications. For instance, is it possible to generate efficiently all (regular) lifts up to isomorphism?
(This is non-trivial since different voltage assignments might lead to isomorphic graphs.) Or, does
there exists a lift with some additional properties?

Concerning \quotient, by Theorem~\ref{thm:lubiw}, this problem is \cNP-complete even for the fixed $k=2$.
(We ask for existence of a half-quotient $H$ of $G$ which is equivalent to existence of a
fixed-point free involution in $\Aut(G)$.) This hardness reduction can be easily generalized for
every fixed even $k$, but the complexity remains open for odd values of $k$.

The reduction of Theorem~\ref{thm:lubiw} is from 3-satisfiability, each variable is represented by a
variable gadget which is an even cycle attached to the rest of the graph.  Each cycle has two
possible regular quotients, either the cycle of half length (obtained by the $180^\circ$ rotation), or
the path of half length with attached half-edges (obtained by a reflection through opposite edges),
corresponding to true and false values, respectively.  These variable gadgets are attached to clause
gadgets, and a quotient of a clause gadget can be constructed if and only if at least one literal of
the clause is satisfied.  This reduction does not imply \cNP-completeness for the \rcover problem
since the input also gives a graph $H$, so one can decode the assignment of the variables from it.

\heading{$\boldsymbol k$-Fold Covering.}
To simplify the \rcover problem, instead of fixing $H$, we can fix the ratio $k = |G|/|H|$. (When
$G$ covers $H$, then $k$ is an integer.) We get the following two problems for general and regular
graph covers, respectively:

\computationproblem
{\foldbracketrcover{k}}
{Connected graphs $G$ and $H$ such that $|G| = k|H|$.}
{Does $G$ (regularly) cover $H$?}
{7.5cm}

For $k=1$, both problems are equivalent to \gi.  Bodlaender~\cite{bodlaender} proved that the
\foldcover{k} problem is \cGI-hard for every fixed $k$ (meaning that \gi can be reduced to it). The
same reduction also works for \foldrcover{k}, see Lemma~\ref{lem:gi_hardness}. Chaplick et
al.~\cite{3fold_cover_npc} proved \cNP-completeness of \foldcover{3} and their reduction can be
easily modified for all $k > 3$.  The complexities of \foldcover{2} and \foldrcover{k} for all $k
\ge 2$ are open and very interesting. We note that for $k=2$, every covering is a regular covering,
so the problems \foldrcover{2} and \foldcover{2} are identical, and \cNP-hardness of
\foldcover{2} would imply \cNP-hardness for \rcover as well. On the other hand, if \foldrcover{k} is
not \cNP-complete for any value $k$, the \foldrcover{k} problems would be natural generalizations of
\gi.

\subsection{Three Properties}

Let $\calC$ be a class of connected multigraphs. By $\calC / \Gamma$ we denote the class of all
regular quotients of graphs of $\calC$ (note that $\calC \subseteq \calC / \Gamma$). For instance,
when $\calC$ is the class of planar graphs, then the class $\calC / \Gamma$ is, by Negami
Theorem~\cite{negami}, the class of projective planar graphs. We define the following three
properties of $\calC$, for formal definitions see Section~\ref{sec:preliminaries}:
\begin{packed_head_enum}{(P1)}
\item[(P1)] The classes $\calC$ and $\calC / \Gamma$ are closed under taking subgraphs and under
replacing connected components attached to 2-cuts by edges. 
\item[(P2)] For a 3-connected graph $G \in \calC$, all semiregular subgroups $\Gamma$ of $\Aut(G)$
can be computed in polynomial time. Here by semiregularity, we mean that the action of $\Gamma$ has
no non-trivial stabilizers of the vertices.
\item[(P3)] Let $G$ and $H$ be 3-connected graphs of $\calC / \Gamma$, possibly with colored and
directed edges, and the vertices of $G$ be equipped with lists. We can decide \lgi of $G$ and $H$ in
polynomial time. (Where the list-compatible isomorphism respects orientations and colors of edges.)
\end{packed_head_enum}
As we prove in Lemma~\ref{lem:planar_graph_properties}, these three properties are tailored for the
class of planar graphs. (The proof of the property (P3) is non-trivial, following from~\cite{kkz}.)
The main reason to state (P1) to (P3) is explicitely to make clear which properties of planar graphs
are necessary for our algorithm.

Since \lgi is \cNP-complete in general, we also use the restricted version with only \cgi to
highlight places where \lgi can be avoided:
\begin{packed_head_enum}{(P$3^*$)}
\item[(P$3^*$)] Let $G$ and $H$ be 3-connected graphs of $\calC / \Gamma$, possibly with colored and
directed edges, and the vertices of $G$ and $H$ are colored. We can decide \cgi of $G$ and $H$ in
polynomial time.
\end{packed_head_enum}

\subsection{The Meta-algorithm}

This paper studies complexity of regular covering testing, based on our structural results described
in~\cite{fkkn16}. We establish the following algorithmic result:

\begin{theorem} \label{thm:metaalgorithm}
Let $\calC$ be a class of graphs satisfying (P1) to (P3). There exists an \cFPT\ algorithm for
\rcover for $\calC$-inputs $G$ in time $\O^*(2^{\be(H)/2})$, where $\be(H)$ is the number of edges
of $H$.
\end{theorem}

Since the assumptions (P1) to (P3) are satisfied for planar graphs
(Lemma~\ref{lem:planar_graph_properties}), we get Theorem~\ref{thm:planar_rcover}.  Notice that if the input
graph $G$ is 3-connected, using our assumptions the \rcover problem can be trivially solved, by
enumerating all its regular quotients and testing graph isomorphism with $H$.
Babai~\cite{babai1975automorphism} proved that to solve graph isomorphism, it is sufficient to solve
graph isomorphism for 3-connected graphs. We wanted to generalize this result to regular covers, but
handling 2-cuts is very complicated and we need the assumptions (P2) and (P3).

We process the graph $G$ by a series of \emph{reductions}, replacing parts of the graph by edges,
essentially forgetting details of the graph. We end-up with a primitive graph which is either
3-connected, or very simple (a cycle or $K_2$).  This very natural idea of reductions was first
introduced in a seminal paper of Trakhtenbrot~\cite{trakhtenbrot} and further extended
in~\cite{tutte_connectivity,hopcroft_tarjan_dividing,cunnigham_edmonds,walsh,babai1975automorphism}.
The main difference is that these papers apply the reduction only to 2-connected graphs, but
in~\cite{fkkn16}, we also reduce parts separated by 1-cuts.  The reason is that a regular quotient
of a 2-connected graph might be only 1-connected, see Sections~\ref{sec:block_trees}
and~\ref{sec:algorithm_overview}.  Also, we prove in~\cite{fkkn16} that no essential information of
semiregular actions is lost during reductions.

In~\cite{fkkn16}, we describe how regular covering behaves with respect to vertex 1-cuts and 2-cuts.
Concerning 1-cuts, regular covering behaves non-trivially only on the central block of $G$, so they
are easy to deal with. But regular covering can behave highly complex on 2-cuts. In this paper, we
build an algorithm based on these structural results of~\cite{fkkn16}.  When the reductions reach a
3-connected graph, the natural next step is to compute all its quotients; there are polynomially
many of them according to (P2). 

What remains is the most difficult part: To test for each quotient whether it corresponds to $H$
after unrolling the reductions which is called \emph{expanding}. The issue is that there may be
exponentially many different ways to expand the graph, all described in~\cite{fkkn16}.  Therefore,
we have to test in a clever way whether it is possible to reach $H$. Our algorithm consists of
several subroutines, most of which we can perform in polynomial time.  Only one subroutine (finding
a certain ``generalized matching'') we have not been able to solve in polynomial time.

This slow subroutine can be avoided in some cases:

\begin{corollary} \label{cor:simple_cases}
If $G$ is a 3-connected graph, if $H$ is a 2-connected graph, or if $k=|G|/|H|$ is odd, then the
meta-algorithm of Theorem~\ref{thm:metaalgorithm} can be modified to run in polynomial time.
\end{corollary}

\begin{corollary} \label{cor:listing_all_quotients}
Let $\calC$ be a class of graphs satisfying (P1), (P2), and (P$3^*$). There exists an algorithm listing for 
$\calC$-inputs $G$ all their regular quotients, with a polynomial-time delay.
\end{corollary}

Theorem~\ref{thm:lubiw} implies that to solve the \rcover\ problem in general, one has to work with both
graphs $G$ and $H$ from the beginning. Our algorithm starts only with $G$ and tries to match its
quotients to $H$ only in the end.  

\heading{Outline.} In Section~\ref{sec:preliminaries}, we introduce the formal notation used in this
paper. In Section~\ref{sec:structural_results}, we state key structural properties of atoms,
reductions and expansions from~\cite{fkkn16}.  In Section~\ref{sec:algorithm}, we use them to design
the meta-algorithm of Theorem~\ref{thm:metaalgorithm}. In Section~\ref{sec:star_atoms}, we describe
more details concerning the only slow subroutine of the meta-algorithm.  Finally, in
Section~\ref{sec:planar_graphs} we show that the class of planar graphs satisfies (P1) to (P3), thus
proving Theorem~\ref{thm:planar_rcover}. In Conclusions, we describe open problems and possible
extensions of our results.

\section{Definitions and Preliminaries} \label{sec:preliminaries}

In this paper, we work with an extended model of graph which is formally described in~\cite{fkkn16}.
A multigraph $G$ is a pair $(\bV(G),\bE(G))$ where $\bV(G)$ is a set of vertices and $\bE(G)$ is a
multiset of edges. We denote $|\bV(G)|$ by $\bv(G) = |G|$ and $|\bE(G)|$ by $\be(G)$.  The graph can
possibly contain parallel edges and loops, and each loop at $u$ is incident twice with the vertex
$u$.  Each edge $e = uv$ gives rise to two half-edges, one attached to $u$ and the other to $v$. We
denote by $\bH(G)$ the collection of all half-edges. We denote $|\bH(G)|$ by $\bh(G)$ and clearly
$\bh(G) = 2\be(G)$.  As quotients, we sometime obtain graphs containing \emph{(standalone)
half-edges} (missing the opposite half-edges).\footnote{Half-edges are sometimes also called darts
or arcs while half-edges with free-ends are called semiedges.} Also, in the reductions, we obtain
\emph{pendant edges}, each consisting of two half-edges, one attached to some vertex $u$, the other
attached to no vertex.

Unless the graph is $K_2$, we remove all vertices of degree 1 while keeping both half-edges.
Assuming that the original graph contains no pendant edges, this removal does not change the
automorphism group and existence of regular covering projections from $G$ to $H$ (when the removal
is applied on both $G$ and $H$).  A pendant edge attached to $v$ is called a \emph{single pendant
edge} if it is the only pendant edge attached to $v$. Most graphs in this paper are assumed to be
connected.

We consider graphs with colored edges and also with three different edge types (directed edges,
undirected edges and a special type called halvable edges).  It might seem strange to consider such
general objects. But when we apply reductions, we replace parts of the graph by edges and the colors
encode isomorphism classes of replaced parts. This allows the algorithm to work with smaller reduced
graphs while preserving important parts of the structure of the original large graph. So even if the
input graphs $G$ and $H$ are simple, more complicated multigraphs are naturally constructed.

We denote groups by capital Greek letters as for instance $\Gamma$. We use $\gS_n$, $\gC_n$, $\gD_n$
and $\gA_n$ to denote symmetric groups, cyclic groups, dihedral groups and alternating groups,
respectively.

\subsection{Automorphisms and Groups}

We state the definitions in a very general setting of multigraphs and half-edges. An
\emph{automorphism} $\pi$ is fully described by a permutation $\pi_h : \bH(G) \to \bH(G)$ preserving
edges and incidences between half-edges and vertices. Thus, $\pi_h$ induces two permutations $\pi_v
: \bV(G) \to \bV(G)$ and $\pi_e : \bE(G) \to \bE(G)$ connected together by the very natural property
$\pi_e(uv) = \pi_v(u) \pi_v(v)$ for every $uv \in \bE(G)$.  In most of situations, we omit subscripts
and simply use $\pi(u)$ or $\pi(uv)$. In addition, we require that an automorphism preserves colors,
edge types and orientation of directed edges.

\heading{Automorphism Groups.}
Let $\Aut(G)$ be the group of all automorphisms of $G$. The \emph{orbit} $[v]$ of a vertex $v \in
\bV(G)$ in the action of $\Gamma \le \Aut(G)$ is the set of all vertices $\{\pi(v) \mid \pi \in
\Gamma \le \Aut(G)\}$, and the orbit $[e]$ of an edge $e \in \bE(G)$ is defined similarly as
$\{\pi(e) \mid \pi \in \Gamma \le \Aut(G)\}$. The \emph{stabilizer} $\Gamma_x$ of $x$ is the
subgroup of all automorphisms which fix $x$.  An action is called \emph{semiregular} if it has no
non-trivial (i.e., non-identity) stabilizers of both vertices and half-edges. Further, we require
the stabilizer of an edge in a semiregular action to be trivial, unless it is a halvable edge, when
it may contain an involution transposing the two half-edges.  We say that a group is
\emph{semiregular} if the associated action is semiregular.  More information on permutation groups
can be found in~\cite{rotman}. 

\subsection{Coverings}

\begin{figure}[b!]
\centering
\includegraphics{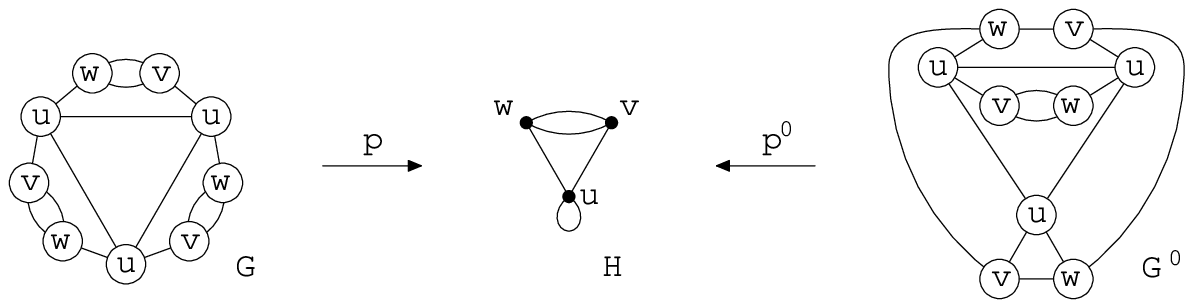}
\caption{Two covers of $H$. The projections $p_v$ and $p'_v$ are written inside of the vertices, and
the projections $p_e$ and $p'_e$ are omitted. Notice that each loop is realized by having two
neighbors labeled the same, and parallel edges are realized by having multiple neighbors labeled the
same. Also covering projections preserve degrees.}
\label{fig:cover_examples}
\end{figure}

A graph $G$ \emph{covers} a graph $H$ (or $G$ is a \emph{cover} of $H$) if there exists a locally
bijective homomorphism $p$ called a \emph{covering projection}.  A homomorphism $p$ from $G$ to $H$
is given by a mapping $p_h : \bH(G) \to \bH(H)$ preserving edges and incidences between half-edges and
vertices. It induces two mappings $p_v : \bV(G) \to \bV(H)$ and $p_e : \bE(G) \to \bE(H)$ such that $p_e(uv)
= p_v(u)p_v(v)$ for every $uv \in \bE(G)$.  The property to be local bijective states that for every
vertex $u \in \bV(G)$ the mapping $p_h$ restricted to the half-edges incident with $u$ is a bijection.
Figure~\ref{fig:cover_examples} contains two examples of graph covers. We mostly omit
subscripts and just write $p(u)$ or $p(e)$.

A \emph{fiber} over a vertex $v \in \bV(H)$ is the set $p^{-1}(v)$, i.e., the set of all vertices $\bV(G)$
that are mapped to $v$, and similarly for fibers over half-edges. From the standard assumption that
both $G$ and $H$ are connected, it follows that all fibers of $p$ are of the same size. In other
words, $|G| = k|H|$ for some $k \in \mathbb N$, which is the size of each fiber, and we say that $G$
is a \emph{$k$-fold cover} of $H$.

\heading{Regular Coverings.} We are going to consider coverings which are highly symmetrical, called
\emph{regular coverings}. For example, in Fig.~\ref{fig:cover_examples}, the covering $p$ is more
symmetric than $p'$. Let $\Gamma$ be a semiregular subgroup of $\Aut(G)$. It defines a graph $G /
\Gamma$ called a \emph{(regular) quotient} of $G$ as follows: The vertices
of $G / \Gamma$ are the orbits of the action of $\Gamma$ on $\bV(G)$, the half-edges of $G / \Gamma$ are
the orbits of $\Gamma$ on $\bH(G)$.  A vertex-orbit $[v]$ is incident with a
half-edge-orbit $[h]$ if and only if the vertices of $[v]$ are incident with the half-edges of
$[h]$. (Because the action of $\Gamma$ is semiregular, each vertex of $[v]$ is incident with exactly
one half-edge of $[h]$, so this is well defined.)

We naturally construct $p : G \to G / \Gamma$ by mapping the vertices to its vertex-orbits and
half-edges to its half-edge-orbits, and it is a $|\Gamma|$-fold regular covering. Concerning an edge
$e \in \bE(G)$, it is mapped to an edge of $G / \Gamma$ if the two half-edges belong to different
half-edge-orbits of $\Gamma$. If both half-edges belong to the same half-edge-orbits, it corresponds
to a standalone half-edge of $G / \Gamma$.

\begin{figure}[t!]
\centering
\includegraphics{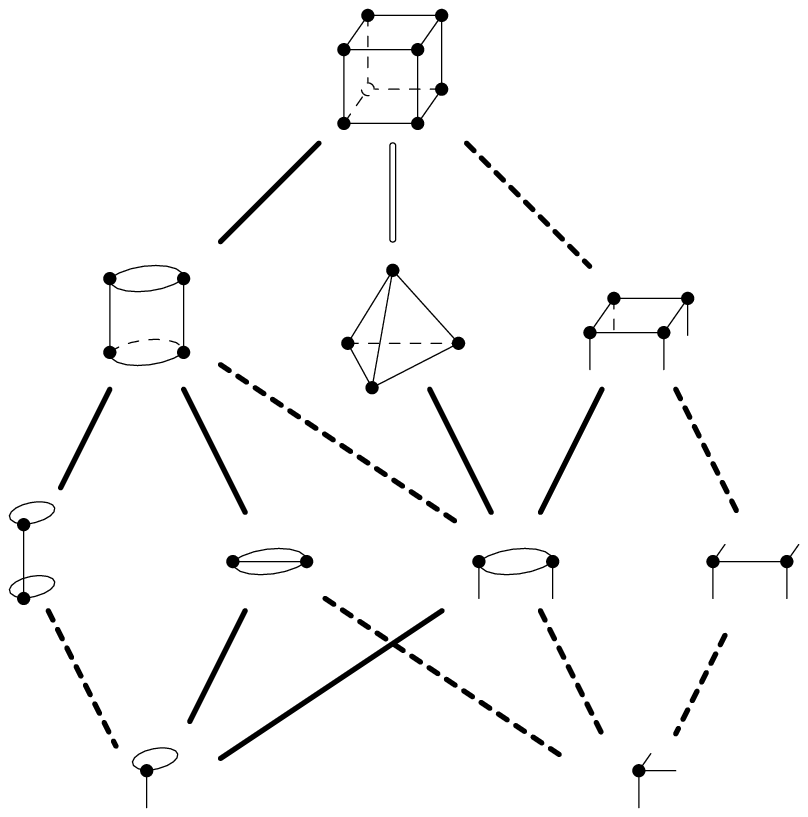}
\caption{The Hasse diagram of all quotients of the cube graph depicted in a geometric way. When
semiregular actions fix edges, the quotients contain half-edges. The quotients connected by bold
edges are obtained by 180 degree rotations. The quotients connected by dashed edges are obtained by
reflections. The tetrahedron is obtained by the antipodal symmetry of the cube, and its quotient is
obtained by a 180 degree rotation with the axis going through the centers of two non-incident edges
of the tetrahedron.}
\label{fig:quotients_of_cube}
\end{figure}

For the graphs $G$ and $H$ of Fig.~\ref{fig:cover_examples}, we get $H \cong G / \Gamma$ for $\Gamma
\cong \gC_3$ which ``rotates the outer cycle by step three'', while $p'$ is not a regular covering.
As a further example, Fig.~\ref{fig:quotients_of_cube} geometrically depicts all regular quotients
of the cube graph.

\subsection{Complexity of Regular Graph Coverings} \label{sec:complexity_properties}

We establish fundamental complexity properties of regular covering. Our goal is to highlight
similarities with the graph isomorphism problem.

\heading{Belonging to NP.} The \cover\ problem clearly belongs to \cNP\ since one can just test in
polynomial time whether a given mapping is a locally bijective homomorphism. Not so obviously, the
same holds for \rcover:

\begin{lemma} \label{lem:rcover_in_np}
The \rcover problem belongs to\/ \cNP.
\end{lemma}

\begin{proof}
By definition, $G$ regularly covers $H$ if and only if there exists a semiregular subgroup $\Gamma$
of $\Aut(G)$ such that $G / \Gamma \cong H$. As a certificate, we give $k$ permutations, one for
each element of $\Gamma$, and an isomorphism between $G / \Gamma$ and $H$. We check that these
permutations define a group $\Gamma$ acting semiregularly on $G$.  The given isomorphism allows to
check that the constructed $G / \Gamma$ is isomorphic to $H$.  Clearly, this certificate is
polynomially large and can be verified in polynomial time.
\end{proof}

One can prove even a stronger result:

\begin{lemma} \label{lem:testing_regularity}
For a mapping $p : G \to H$, we can test whether it is a regular covering projection in polynomial time.
\end{lemma}

\begin{proof}
Testing whether $p$ is a covering projection can clearly be done in polynomial time. It remains to test
regularity. Choose an arbitrary spanning tree $T$ of $H$. Since $p$ is a covering, then $p^{-1}(T)$
is a disjoint union of $k$ isomorphic copies $T_1,\dots,T_k$ of $T$. We number the vertices of the
fibers according to the spanning trees, i.e., $p^{-1}(v) = \{v_1,\dots,v_k\}$ such that $v_i \in
T_i$. This induces a numbering of the half-edges of each fiber over a half-edge of $\bH(H)$, following
the incidences between half-edges and vertices.  For every half-edge $h \notin \bH(T)$, we define the
permutation $\sigma_h$ of $\{1,\dots,k\}$ taking $i$ to $j$ if there is a half-edge $h' \in p^{-1}(h)$
incident with a vertex of $T_i$ and paired with a half-edge incident with a vertex of $T_j$.

Let $\Theta$ be the group generated by all $\sigma_h$, where $h \notin \bH(T)$.  We assume that $G$ is
connected. By Orbit-Stabilizer Theorem, we have $|\Theta| = |\Theta_v| \cdot |[v]|$, and from the
connectivity, it follows that $|[v]| = k$. Therefore, the action of $\Theta$ is regular if and only
if $|\Theta| = k$ which can be checked in polynomial time.
\end{proof}

The constructed permutations $\sigma_h$ associated with $p$ are known in the
literature~\cite{gross_tucker} as \emph{permutation voltage assigments} associated with $p$.

\heading{GI-hardness.}
When $k = |G|/|H| = 1$, the problem \rcover exactly corresponds to \gi.  Let \cGI be the class
of decision problems polynomial-time reducible to \gi.  Bodlaender~\cite{bodlaender} proved
the following for general covers (and his reduction works for regular covers as well):

\begin{lemma} \label{lem:gi_hardness}
For every fixed $k$, the \foldcover{k} and \foldrcover{k} problems are\/ \cGI-hard.
\end{lemma}

\begin{figure}[t!]
\centering
\includegraphics{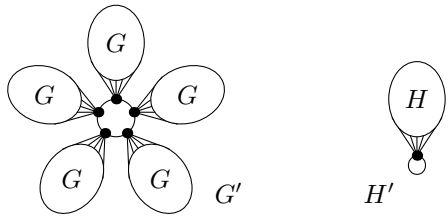}
\caption{The graph $G'$ is constructed by $k$ copies of $G$ with attached universal vertices
connected into a cycle while $H'$ is constructed by attaching a universal vertex with a loop to
$H$.}
\label{fig:gi_hardness}
\end{figure}

\begin{proof}
For input graphs $G$ and $H$ of the graph isomorphism problem, we construct the graphs $G'$ and $H'$
depicted in Fig.~\ref{fig:gi_hardness}. The reduction works since the universal vertices in $G'$ must be
mapped to the universal vertex in $H'$ (since covering projection preserves degrees). Therefore, $G'$
(regularly) covers $H'$ if and only if $G \cong H$.
\end{proof}

\section{Atoms, Reduction and Expansion} \label{sec:structural_results}

In this section, we state definitions and structural results from~\cite{fkkn16}, describing
behaviour of regular graph covers with respect to 1-cuts and 2-cuts in $G$.  We also prove new
results concerning computational complexity of these techniques.  In Section~\ref{sec:block_trees},
we introduce block-trees and describe behaviour of regular covering with respect to $1$-cuts. In
Section~\ref{sec:atoms}, we introduce atoms which are inclusion-minimal parts of $G$ with respect to
$1$-cuts and $2$-cuts. In Section~\ref{sec:reduction}, we describe the reduction which replaces
atoms by colored edges, preserving the essential structure of $G$. In Section~\ref{sec:expansion},
we consider quotients of reduced graphs and revert the reductions in them by expansions.

\subsection{Block-trees and Central Blocks} \label{sec:block_trees}

\begin{figure}[t!]
\centering
\includegraphics{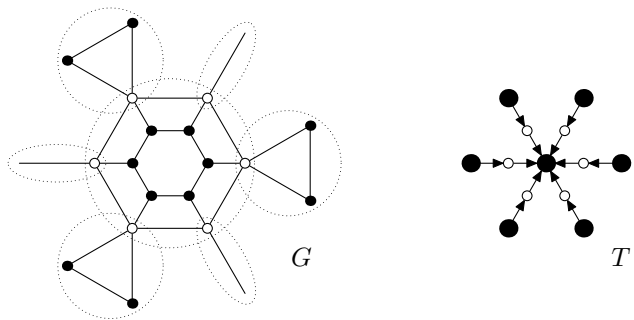}
\caption{On the left, an example graph $G$ with denoted blocks. On the right, the corresponding
block-tree $T$ is depicted, rooted at the central block. The white vertices correspond to the
articulations and the big black vertices correspond to the blocks.}
\label{fig:example_of_block_tree}
\end{figure}

The \emph{block-tree} $T$ of $G$ is defined as follows. Consider all articulations in $G$ and all
maximal $2$-connected subgraphs which we call \emph{blocks} (with bridge-edges and pendant edges
also counted as blocks).  The block-tree $T$ is the incidence graph between the articulations and
the blocks. For an example, see Fig.~\ref{fig:example_of_block_tree}.

\heading{The Central Block.} Recall that for a tree, its \emph{center} is either the central vertex
or the central pair of vertices of a longest path, depending on the parity of its length. Every
automorphism of a tree preserves its center.

\begin{lemma}[\cite{fkkn16}, Lemma 2.1] \label{lem:central_block}
If $G$ has a non-trivial semiregular automorphism, then $G$ has a central block.
\end{lemma}

In the following, we shall assume that $T$ contains a central block $C$. We orient the edges of the
block-tree $T$ towards the central block; so the block-tree becomes rooted. A \emph{(rooted) subtree} of the
block-tree is defined by any vertex different from the central block acting as \emph{root} and by
all its descendants. Let $u$ be an articulation contained in $C$. By $T_u$ we denote the subtree of
$T$ defined by $u$ and all its predecessors, and let $G_u$ be the graph induced by all vertices of
the blocks of $T_u$. All semiregular subgroups of $\Aut(G)$ act non-trivially and faithfully only on
$C$:

\begin{lemma}[\cite{fkkn16}, Lemma 2.2] \label{lem:semiregular_action_on_blocks}
Let $\Gamma$ be a semiregular subgroup of $\Aut(G)$. If $u$ and $v$ are two articulations of the
central block and of the same orbit of $\Gamma$, then $G_u \cong G_v$. Moreover there is a unique
$\pi \in \Gamma$ which maps $G_u$ to $G_v$.
\end{lemma}

In the language of quotients, it means that $G / \Gamma$ consists of $C / \Gamma$ together with
the graphs $G_u$ attached to $C / \Gamma$, one for each orbit of $\Gamma$.

\heading{Why Not Just 2-connected Graphs?}
Since the behaviour of regular covering with respect to 1-cuts in $G$ is very simple, a natural
question follows: why do we not restrict ourselves to 2-connected graphs $G$?
For instance when solving graph isomorphism, it is sufficient to solve graph isomorphism of
2-connected graphs and use it to find isomorphism of block-trees.

This is not possible for regular covering testing.
The issue is that the quotient $C / \Gamma$ might not be 2-connected, so it may consists of many
blocks and it is not easy to locate it in $H$. When $H$ contains a subtree of blocks isomorphic to
$G_u$, it may correspond to $G_u$, or it may correspond to a quotient of a subgraph of $C / \Gamma$,
together with some other $G_v$ attached. We use dynamic programming to deal with this in
the meta-algorithm, see Section~\ref{sec:algorithm_overview}. Therefore, we have to define 3-connected
reduction for 1-cuts in $G$ as well, unlike
in~\cite{trakhtenbrot,hopcroft_tarjan_dividing,cunnigham_edmonds,walsh,babai1975automorphism}.

\subsection{Atoms} \label{sec:atoms}

Suppose that $B$ is a block of $G$, in particular $B$ is 2-connected. Two vertices $u$ and $v$ form a
\emph{2-cut} $U = \{u,v\}$ if $B \setminus U$ is disconnected. We say that a 2-cut $U$ is
\emph{non-trivial} if $\deg(u) \ge 3$ and $\deg(v) \ge 3$ in $B$.

Atoms are inclusion-minimal subgraphs with respect to 1-cuts and 2-cuts in $G$.  We first define a
set $\calP$ of subgraphs of $G$ called \emph{parts} which are candidates for atoms:
\begin{packed_itemize}
\item A \emph{block part} is a subgraph non-isomorphic to a pendant edge induced by the blocks of a
subtree of the block-tree.
\item A \emph{proper part} is a subgraph $S$ of $G$ defined by a non-trivial 2-cut $U$ of a block
$B$. The subgraph $S$ consists of a connected component $K$ of $G \setminus U$ together with $u$ and
$v$ and all edges between $\{u,v\}$ and $K$. In addition, we require that $S$ does not contain the
central block; so it only contains some blocks of the subtree of the block-tree rooted at $B$.
\item A \emph{dipole part} is any dipole defined as follows.  Let $u$ and $v$ be two distinct
vertices of degree at least three joined by at least two parallel edges. Then the subgraph induced
by $u$ and $v$ is called a \emph{dipole}.
\end{packed_itemize}

\begin{figure}[t!]
\centering
\includegraphics{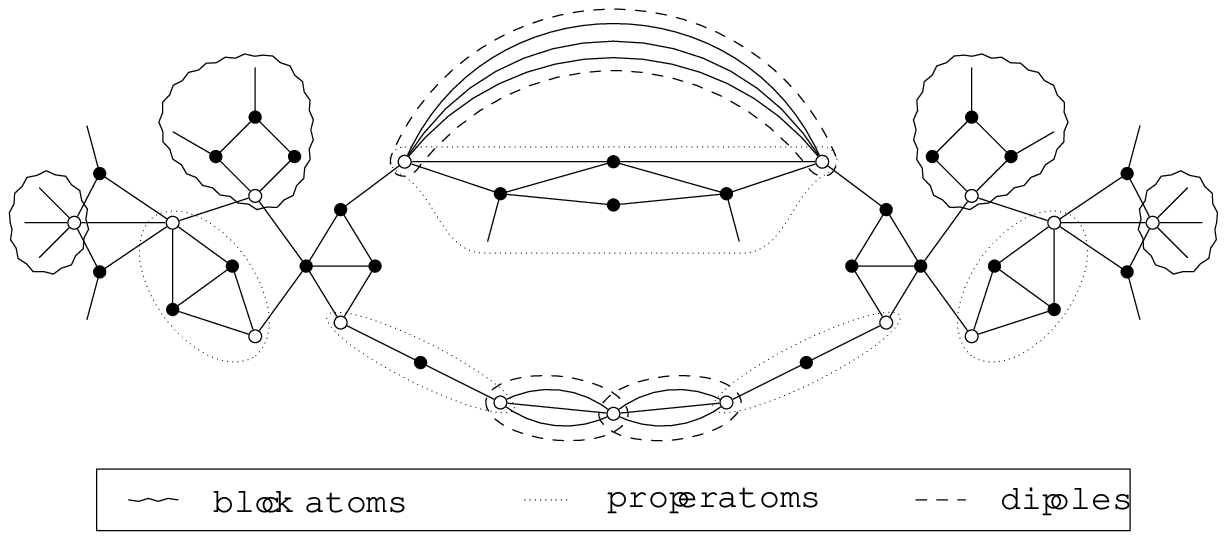}
\caption{An example of a graph with denoted atoms. The white vertices belong to the boundary of some
atom, possibly several of them.}
\label{fig:atoms_examples}
\end{figure}

\noindent The inclusion-minimal elements of $\calP$ are called \emph{atoms}. We distinguish
\emph{block atoms}, \emph{proper atoms} and \emph{dipoles} according to the type of the defining
part.  Block atoms are either stars of pendant edges called \emph{star block atoms}, or pendant blocks
possibly with single pendant edges attached to them called \emph{non-star block atoms}. Each proper
atom is a subgraph of a block, together with some single pendant edges attached to it. A dipole part
is by definition always inclusion-minimal, and therefore it is an atom. For an example, see
Fig.~\ref{fig:atoms_examples}.

We use the topological notation to denote the \emph{boundary} $\bo A$  and the \emph{interior} $\int A$
of an atom $A$.  If $A$ is a dipole, we set $\bo A = \bV(A)$. If $A$ is a proper or block atom, we put
$\bo A$ equal to the set of vertices of $A$ which are incident with an edge not contained in $A$. For
the interior, we use the standard topological definition $\int A = A \setminus \bo A$ where we only
remove the vertices $\bo A$, the edges adjacent to $\bo A$ are kept in $\int A$.
Single pendant edges of $A$ are always attached to $\int A$.

Note that $|\bo A| = 1$ for a block atom $A$, and $|\bo A| = 2$ for a proper atom or dipole $A$.
The interior of a dipole is a set of free edges.  For a proper atom $A$, the vertices of
$\bo A$ are exactly the vertices $\{u,v\}$ of the non-trivial 2-cut used in the definition of proper
parts, and they are never adjacent in $A$.

\begin{lemma}[\cite{fkkn16}, Lemma 3.3] \label{lem:nonintersecting_atoms}
Let $A$ and $A'$ be two different atoms. Then $A \cap A' = \bo A \cap \bo A'$.
\end{lemma}

\begin{lemma}[\cite{fkkn16}, Lemma 3.8] \label{lem:atom_automorphisms}
Let $A$ be an atom and let $\pi \in \Aut(G)$.
\begin{packed_enum}
\item[(a)] The image $\pi(A)$ is an atom isomorphic to $A$. Moreover $\pi(\bo A) = \bo \pi(A)$ and
$\pi(\int A) = \int \pi(A)$, where $\int \pi(A)$ denotes the interior of $\pi(A)$.
\item[(b)] If $\pi(A) \ne A$, then $\pi(\int A) \cap \int A = \emptyset$.
\item[(c)] If $\pi(A) \ne A$, then $\pi(A) \cap A = \bo A \cap \bo \pi(A)$.
\end{packed_enum}
\end{lemma}

\begin{figure}[b!]
\centering
\includegraphics{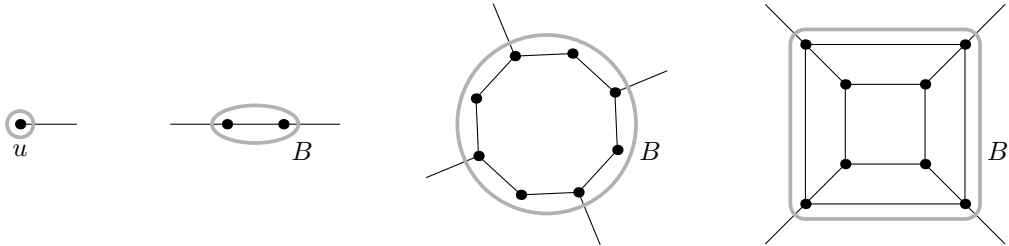}
\caption{A primitive graph with a central block is either $K_2$, $C_n$, or a 3-connected graph, in
all three cases with possible single pendant edges attached to it.}
\label{fig:primitive_graphs}
\end{figure}

\heading{Primitive Graphs.} 
A graph is called \emph{primitive} if it contains no atoms.  The following lemma characterizing
primitive graphs can be alternatively obtained from the well-known theorem by
Trakhtenbrot~\cite{trakhtenbrot}, see Fig.~\ref{fig:primitive_graphs} for examples.\footnote{We
consider $K_1$ with an attached single pendant edge as a graph with a central articulation.}

\begin{lemma}[\cite{fkkn16}, Lemma 3.4] \label{lem:primitive_graphs}
Let $G$ be a primitive graph. If $G$ has a central block, then it is a 3-connected
graph, a cycle $C_n$ for $n \ge 2$, or $K_2$, or can be obtained from the aforementioned graphs by attaching
single pendant edges to at least two vertices. If $G$ has a central articulation, then it is $K_1$,
possible with a single pendant edge attached.
\end{lemma}

\begin{lemma} \label{lem:testing_gi_primitive_graphs}
If primitive graphs $G$ and $G'$ belong to $\calC$ satisfying (P3$\,{}^*\!$), then we can test $G \cong
G'$ in polynomial time.
\end{lemma}

\begin{proof}
In both graphs, we replace single pendant edges with colored vertices. If $G$ and $G'$ are $K_1$, or $K_2$,
the problem is trivial. If they are cycles, we use the standard cycle isomorphism algorithms. If they
are 3-connected, we test $G \cong G'$ using (P3${}^*$).
\end{proof}

\heading{Structure of Atoms.}
We call a graph \emph{essentially 3-connected} if it is a 3-connected graph with possibly single
pendant edges attached to it.  Similarly, a graph is called \emph{essentially a cycle} if it is a
cycle with possibly single pendant edges attached to it. The structure of dipoles and star block
atoms is clear. Non-star block and proper atoms are either almost 3-connected, or very simple:

\begin{lemma}[\cite{fkkn16}, Lemma 3.5] \label{lem:non_star_block_atoms}
Every non-star block atom $A$ is either $K_2$ with an attached single pendant edge, essentially a
cycle, or essentially 3-connected.
\end{lemma}

Let $A$ be a proper atom with $\bo A = \{u,v\}$. We define the \emph{extended proper atom} $A^+$ as
$A$ with the additional edge $uv$.  Notice that the property (P1) ensures that $A^+$ belongs to
the class $\calC$.

\begin{lemma}[\cite{fkkn16}, Lemma 3.6] \label{lem:proper_atoms}
For every proper atoms $A$, the extended proper atom $A^+$ is either essentially a cycle, or
essentially 3-connected.
\end{lemma}

Two atoms $A$ and $A'$ are isomorphic if there exists an isomorphism which maps $\bo A$ to $\bo A'$.

\begin{lemma} \label{lem:testing_gi_atoms}
For every atoms $A$ and $A'$ of a graph $G$ belonging to $\calC$ satisfying (P1) and (P3$\,{}^*\!$), we
can test $A \cong A'$ in polynomial time.
\end{lemma}

\begin{proof}
If both $A$ and $A'$ are dipoles and star block atoms, we can test $A \cong A'$ trivially in
polynomial time. If they are non-star block atoms, by Lemma~\ref{lem:non_star_block_atoms} they are
either $K_2$ with attached single pendant edge, or essentially a cycle, or essentially 3-connected.
The first two possibilities can be solved trivially, so we assume that $A$ and $A'$ are essentially
3-connected. Let $B$ and $B'$ be the 3-connected graph created from $A$ and $A'$ by removing pendant
edges, where existence of pendant edges is coded by colors of $\bV(B)$ and $\bV(B')$, and we further
color $\bo B$ and $\bo B'$ by a special color. We have $A \cong A'$ if and only if there exists a
color-preserving isomorphism between $B$ and $B'$ which can be tested using (P3${}^*$). When $A$ and
$A'$ are proper atoms, we proceed similarly on extended proper atoms, using
Lemma~\ref{lem:proper_atoms}.
\end{proof}

\heading{Symmetry Types of Atoms.}
We distinguish three symmetry types of atoms, and in reductions, we replace atoms by edges carrying
their types. Therefore we work with multigraphs with three edge types: \emph{halvable edges},
\emph{undirected edges} and \emph{directed edges}. We consider only the automorphisms which preserve
these edge types and of course the orientation of directed edges. For an atom $A$, we denote by
$\Aut(A)$ the setwise stabilizer of $\bo A$.

\begin{figure}[b!]
\centering
\includegraphics{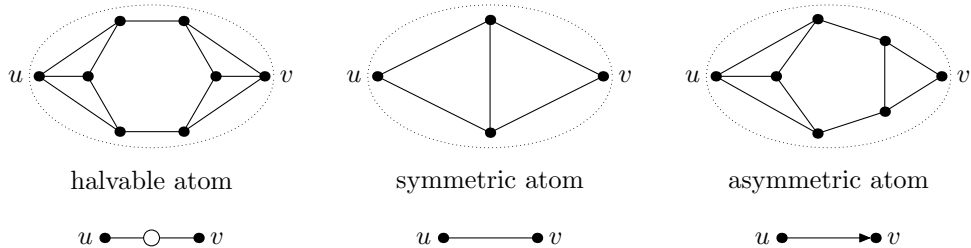}
\caption{The three types of atoms and the corresponding edge types which we use in the reduction. We
denote halvable edges by small circles in the middle.}
\label{fig:types_of_atoms}
\end{figure}

Let $A$ be a proper atom or dipole with $\bo A = \{u,v\}$. We distinguish the following
three symmetry types, depicted in Fig.~\ref{fig:types_of_atoms}:
\begin{packed_itemize}
\item \emph{The halvable atom.} There exists a semiregular involutory automorphism $\tau \in \Aut(A)$
which exchanges $u$ and $v$. More precisely, the automorphism $\tau$ fixes no vertices and no
directed and undirected edges, but some halvable edges may be fixed.
\item \emph{The symmetric atom.} The atom is not halvable, but there exists an automorphism in
$\Aut(A)$ which exchanges $u$ and $v$.
\item \emph{The asymmetric atom.} The atom is neither halvable, nor symmetric.
\end{packed_itemize}
If $A$ is a block atom, then it is by definition symmetric.

\begin{figure}[t!]
\centering
\includegraphics{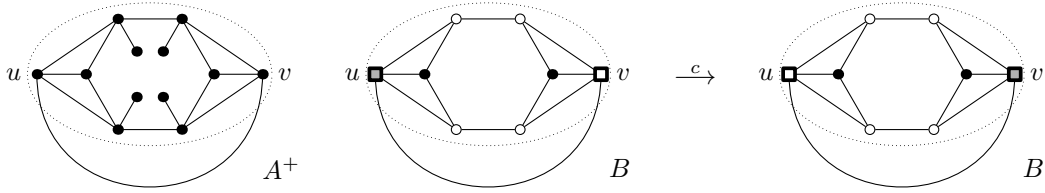}
\caption{For the depicted atom $A$, we test using (P3$\,{}^*\!$) whether $B \protect\coliso B$. In this case yes, so
$A$ is either symmetric, or halvable.}
\label{fig:testing_symmetry}
\end{figure}

\begin{lemma} \label{lem:dipole_symmetry_type}
For a dipole $A$, we can determine its symmetry type in polynomial time.
\end{lemma}

\begin{proof}
The type depends only on the quantity of distinguished types of the parallel edges.  We have
directed edges from $u$ to $v$, directed edges from $v$ to $u$, undirected edges and halvable edges.
We call a dipole \emph{balanced} if the number of directed edges in the both directions is the same.
The dipole is halvable if and only if it is balanced and has an even number of undirected edges. The
dipole is symmetric if and only if it is balanced and has an odd number of undirected edges. The
dipole is asymmetric if and only if it is unbalanced. This clearly can be tested in polynomial time.
\end{proof}

\begin{lemma} \label{lem:proper_atom_symmetry_type}
For a proper atom $A$ of $\calC$ satisfying (P1), (P2), and (P3$\,{}^*\!$), we can determine its symmetry
type in polynomial time.
\end{lemma}

\begin{proof}
Let $\bo A = \{u,v\}$. By Lemma~\ref{lem:proper_atoms}, $A^+$ is either essentially a cycle (which
is easy to deal with), or an essentially 3-connected graph. Let $B$ be the 3-connected graph created
from $A^+$ by removing pendant edges, where existence of pendant edges is coded by colors of $\bV(B)$.
By (P1), both $A^+$ and $B$ belong to $\calC$.  We apply (P3${}^*$) on two copies of $B$. In one copy,
we color $u$ by a special color, and $v$ by another special color. In the other copy, we swap the
colors of $u$ and $v$. Using (P3${}^*$), we check whether there exists a color-preserving
automorphism which exchanges $u$ and $v$; see Fig.~\ref{fig:testing_symmetry}. If not, then $A$ is
asymmetric. If yes, we check whether $A$ is symmetric or halvable.

Using (P2), we generate polynomially many semiregular involutions of order two acting on $B$.  For
each semiregular involution, we check whether it transposes $u$ to $v$, and whether it preserves the
colors of $\bV(B)$ coding pendant edges. If such a semiregular involution exists, then $A$ is halvable,
otherwise it is just symmetric.
\end{proof}

\begin{figure}[b!]
\centering
\includegraphics{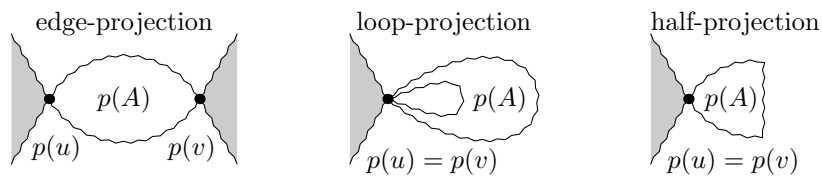}
\caption{How can $p(A)$ look in $G / \Gamma$, for three types of projection.}
\label{fig:projections_of_atoms}
\end{figure}

\heading{Regular Projections and Quotients of Atoms.}
Let $\Gamma$ be a semiregular subgroup of $\Aut(G)$, which defines a regular covering projection $p
: G \to G / \Gamma$. For a proper atom or a dipole $A$ with $\bo A = \{u,v\}$, we get three possible
\emph{types of projection} $p |_A$; see Fig.~\ref{fig:projections_of_atoms}:

\begin{packed_itemize}
\item \emph{An edge-projection.} The atom $A$ is preserved in $G / \Gamma$, meaning $p(A) \cong A$.
Notice that $p(A)$ may just be a subgraph of $G / \Gamma$, not induced. For instance,
it can happen that $p(u)p(v) \in \bE(G / \Gamma)$ while $uv \notin \bE(G)$.
\item \emph{A loop-projection.} The interior $\int A$ is preserved and the vertices $u$ and $v$ are
identified, i.e., $p(\int A) \cong \int A$ and $p(u) = p(v)$.
\item \emph{A half-projection.} There exists an involutory permutation $\pi$ in $\Gamma$ which
exchanges $u$ and $v$ and preserves $A$. The projection $p(A)$ is a halved atom $A$. This can happen
only when $A$ is a halvable atom.  In particular, the covering projection $p$ is a $2k$-fold
covering.
\end{packed_itemize}

\begin{lemma}[\cite{fkkn16}, Lemma 3.9] \label{lem:atom_covering_cases}
For an atom $A$ and a regular covering projection $p$, we have $p |_A$ either an edge-projection, a
loop-projection, or a half-projection. Moreover, for a block atom we have exclusively an edge-projection.
\end{lemma}

So we get three types of quotients $p(A)$ of $A$. For an edge-projection, we call this quotient an
\emph{edge-quotient}, for a loop-projection, we call it a \emph{loop-quotient}, and for a
half-projection, we call it a \emph{half-quotient}. The following lemma allows to say ``the'' edge-
and ``the'' loop-quotient of an atom.

\begin{lemma}[\cite{fkkn16}, Lemma 3.9] \label{lem:unique_quotients}
For every atom $A$, there is the unique edge-quotient and the unique loop-quotient up to isomorphism.
\end{lemma}

\begin{figure}[t!]
\centering
\includegraphics{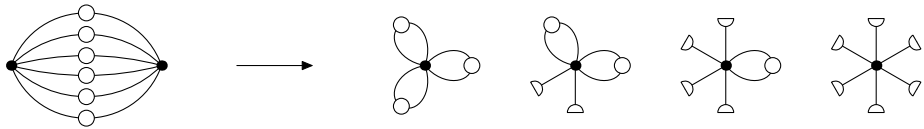}
\caption{Assuming that quotients can contain half-edges, the depicted dipole has four
non-isomorphic half-quotients.}
\label{fig:halfquotients_of_dipole}
\end{figure}

For half-quotients, this uniqueness does not hold. First, an atom $A$ with $\bo A = \{u,v\}$ has to
be halvable to admit a half-quotient. Then each half-quotient is determined by an involutory
automorphism $\tau$ exchanging $u$ and $v$; here $\tau$ is the restriction of $\pi$ from the
definition of a half-projection. First, several different automorphisms $\tau$ may define equivalent
covering projections $p$ of $A$, so $p(A)$ is an isomorphic half-quotient. On the other hand,
different automorphisms $\tau$ may define non-equivalent covering projections $p$ of $A$, so they
give non-isomorphic half-quotients $p(A)$; see Fig.~\ref{fig:halfquotients_of_dipole}. For a proper
atom, we can bound the number of non-isomorphic half-quotients by the number of different
semiregular involutions of 3-connected graphs.

\begin{lemma} \label{lem:proper_atom_quotients}
Let $A$ be a proper atom of $\calC$ satisfying (P1) and (P2). Then there are polynomially many
non-isomorphic half-quotients of $A$ which can be computed in polynomial time.
\end{lemma}

\begin{proof}
By Lemma~\ref{lem:proper_atoms}, $A^+$ is either essentially a cycle (where it holds trivially), or
it is an essentially 3-connected graph. We construct $B^+$ from $A^+$ be replacing pendant edges
with colored vertices, by (P1) both $A^+$ and $B^+$ belong to $\calC$. According to (P2), the number
of different semiregular subgroups of order two is polynomial in the size of $B^+$. Each
half-quotient is defined by one of these semiregular involutions which fixes the edge $uv$,
transposes $u$ and $v$, and preserves colors.
\end{proof}

\subsection{Reduction} \label{sec:reduction}

The reduction produces a \emph{reduction series} of graphs $G = G_0,\dots,G_r$.  It produces graphs
with colored edges and with three edge types: halvable, undirected and directed.  We note that the
results built in Section~\ref{sec:atoms} transfers to colored graphs and colored atoms without any
problems. 

To construct $G_{i+1}$ from $G_i$, we find the collection of all atoms $\calA$ of $G_i$, together
with isomorphism classes such that $A$ and $A'$ belong to the same class if and only if $A \cong
A'$.  To each isomorphism class, we assign one new color not yet used in the graphs $G_0,\dots,G_i$.
We replace the atoms $\calA$ in $G_i$ by edges of colors of the
corresponding isomorphism classes as follows.

\begin{figure}[b!]
\centering
\includegraphics{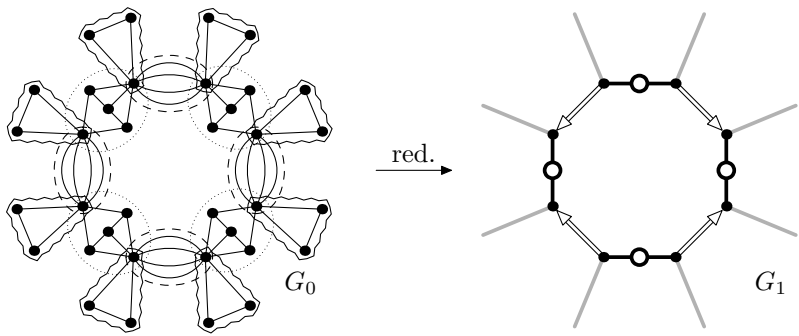}
\caption{On the left, we have a graph $G_0$ with three isomorphism classes of atoms.
The dipoles are halvable, the block atoms are symmetric and the proper atoms are asymmetric.  We
reduce $G_0$ to $G_1$ which is an eight cycle with single pendant edges, with four black halvable
edges replacing the dipoles, eight gray undirected edges replacing the block atoms, and four white
directed edges replacing the proper atoms. The reduction series ends with $G_1$ since it is
primitive.}
\label{fig:example_of_reduction}
\end{figure}

For each block atom $A$ with $\bo A = \{u\}$, we replace it by a pendant edge of
some color based at $u$.  For each proper atom or dipole $A$ with $\bo A = \{u,v\}$, we replace it
by a new edge $uv$ which is halvable/undirected/directed when $A$ is halvable/symmetric/asymmetric,
respectively.  Naturally, for each isomorphism class of asymmetric atom, we consistently choose an
arbitrary orientation of the directed edges replacing these atoms. For an example of the reduction,
see Fig.~\ref{fig:example_of_reduction}.  By Lemma~\ref{lem:nonintersecting_atoms}, the replaced the
interiors of the atoms of $\calA$ are pairwise disjoint, so the reduction is well defined.

\begin{figure}[t!]
\centering
\includegraphics{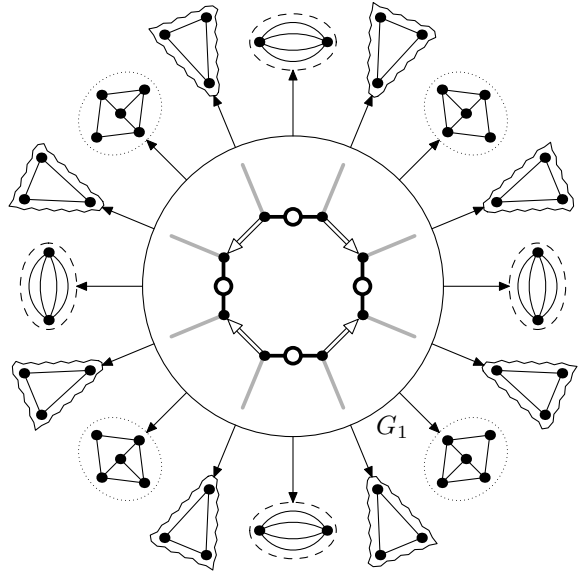}
\caption{The reduction tree for the reduction series in Fig.~\ref{fig:example_of_reduction}. The
root is the primitive graph $G_1$ and each leaf corresponds to one atom of $G_0$.}
\label{fig:reduction_tree}
\end{figure}

The reduction series stops in the step $r$ when $G_r$ is a primitive graph.
For every graph $G$, the reduction series corresponds to the \emph{reduction tree} which is
a rooted tree defined as follows.  The root is the primitive graph $G_r$, and the other nodes are
the atoms obtained during the reductions. If a node contains a colored edge, it has the
corresponding atom as a child. Therefore, the leaves are the atoms of $G_0$, after removing them,
the new leaves are the atoms of $G_1$, and so on. For an example, see Fig.~\ref{fig:reduction_tree}.
It is proved in~\cite{fkkn16,knz} that the reduction series and the reduction tree captures $\Aut(G)$.

\begin{lemma} \label{lem:reduction_series}
If a graph $G$ belongs to $\calC$ satisfying (P1), (P2) and (P3$\,{}^*\!$), then the reductions series
$G = G_0,\dots,G_r$ and the reduction tree can be computed in polynomial time.
\end{lemma}

\begin{proof}
To compute $G_{i+1}$ from $G_i$, we find all atoms $\calA$ in $G_i$ and $\calA'$ in $G'_i$, compute
their isomorphism classes by Lemma~\ref{lem:testing_gi_atoms} and assign new colors to them. By
Lemmas~\ref{lem:dipole_symmetry_type} and~\ref{lem:proper_atom_symmetry_type}, we compute symmetry
types of these atoms. We end up with a primitive graph $G_r$ containing the atoms.  The reduction
tree can be easily constructed and the algorithm runs in polynomial time. 
\end{proof}

\begin{figure}[b!]
\centering
\includegraphics{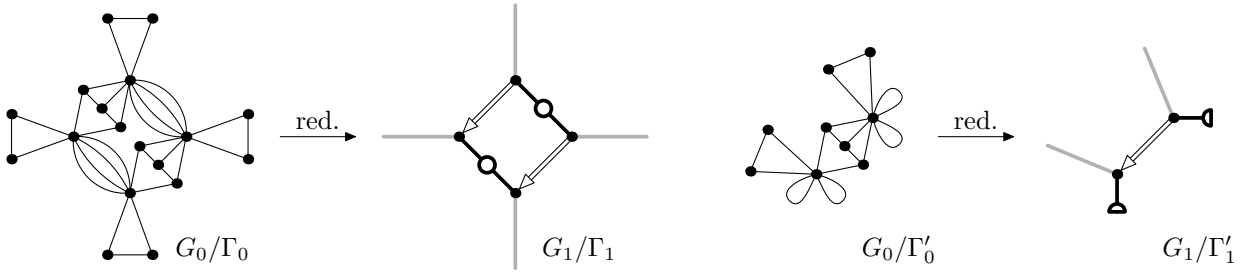}
\caption{An example of two quotients $G_0 / \Gamma_0$ and $G_0 / \Gamma_0'$ of the graph $G_0$ from
Fig.~\ref{fig:example_of_reduction} with the corresponding quotients of the reduced graph $G_1$.}
\label{fig:computed_quotients}
\end{figure}

The following is the approach invented by Babai~\cite{babai1975automorphism}:

\begin{lemma}[\cite{babai1975automorphism}] \label{lem:testing_gi}
If graphs $G$ and $G'$ belong to $\calC$ satisfying (P1) and (P3$\,{}^*\!$), we can test $G \cong G'$ in
polynomial time.
\end{lemma}

\begin{proof}
Using Lemma~\ref{lem:reduction_series}, we simultaneously apply reduction series on both $G$ and
$G'$ in polynomial time, using identical colors for isomorphic atoms in $G_i$ and $G'_i$. (We do not
need to distinguish halvable and symmetric atoms, so (P2) is not needed.) We end up with two
primitive graphs $G_r$ and $G'_r$ and we test their isomorphism using
Lemma~\ref{lem:testing_gi_primitive_graphs}.  Alternatively, we can compute both reduction trees and
apply the standard tree isomorphism of graph-labeled trees.
\end{proof}

In general, the reduction series does not have to preserve the central block, and the atoms $\calA$
of $G_0,\dots,G_{r-1}$ has to be defined with respect to one chosen block which is preserved.  On
the other hand, by Lemma~\ref{lem:central_block}, if the \rcover problem is non-trivial, then $G$
contains the central block which is preserved by the reduction series:

\begin{lemma}[\cite{fkkn16}, Lemma 4.1] \label{lem:preserved_center}
Let $G$ admit a non-trivial semiregular automorphism $\pi$. Then each $G_{i+1}$ has a central block
which is obtained from the central block of $G_i$ by replacing its atoms by colored edges.
\end{lemma}

\begin{figure}[t!]
\centering
\includegraphics{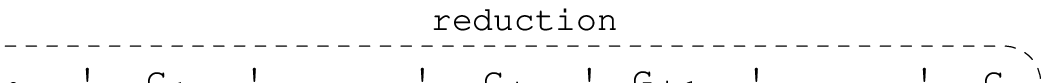}
\caption{The graph $H_{i+1}$ is constructed from $H_i$ by replacing the projections of atoms in
$H_i$ by the corresponding projections of the edges replacing the atoms.}
\label{fig:reduction_replacement_diagram}
\end{figure}

\heading{Quotient Reduction.}
Let $G_0,\dots,G_r$ be the reduction series of $G$ and let $\Gamma_0$ be a semiregular subgroup of
$\Aut(G_0)$. By Lemma~\ref{lem:atom_automorphisms}, we argue that $\Gamma_0$ uniquely determines
semiregular subgroups $\Gamma_1,\dots,\Gamma_r$ of $\Aut(G_1),\dots,\Aut(G_r)$.  Each element $\pi
\in \Gamma_i$ somehow permutes atoms $\calA$ of $G_i$ and somehow permutes the rest of $G_i$. The
reduction constructs $G_{i+1}$ from $G_i$ by replacing $\calA$ with colored edges.  Therefore, $\pi'
\in \Gamma_{i+1}$ corresponding to $\pi$ can be defined in the following way. It permutes the
colored edges in the same way as $\pi$ permutes the respective atoms of $\calA$, while $\pi'$ is
equal to $\pi$ on the rest of the graph. The mapping $\pi \mapsto \pi'$ defines the \emph{reduction
epimorphism} $\Phi_i : \Aut(G_i) \to \Aut(G_{i+1})$; see~\cite{fkkn16}, Proposition 4.1.

Let $H_i = G_i / \Gamma_i$ be the quotients with preserved colors and types of edges, and let
$p_i$ be the corresponding covering projection from $G_i$ to $H_i$. Recall that $H_i$ can contain
edges, loops and half-edges; depending on the action of $\Gamma_i$ on the half-edges corresponding
to the edges of $G_i$. We investigate relation between $H_i = G_i / \Gamma_i$ and $H_{i+1} = G_{i+1}
/ \Gamma_{i+1}$. 

Let $A$ be an atom of $G_i$ represented by a colored edge $e$ in $G_{i+1}$. By
Lemma~\ref{lem:atom_covering_cases}, $p_i |_A$ can have three types of projections. It is easy to
see that $p_{i+1}(e)$ corresponds to an edge (for a block atom, to a pendant edge) for the
edge-projection, to a loop for the loop-projection and to a half-edge for a half-projection. This
explains the names of the quotients $p_i(A)$ as the edge-quotient, the loop-quotient and a
half-quotient.  See Fig.~\ref{fig:computed_quotients} for examples and see the diagram in
Fig.~\ref{fig:reduction_replacement_diagram}.

\subsection{Expansion} \label{sec:expansion}

\begin{figure}[b!]
\centering
\includegraphics{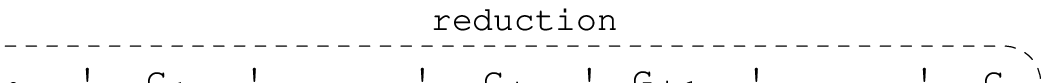}
\caption{The expansion constructs the graph $H_i$ from $H_{i+1}$ by replacing the edges, loops and
half-edges corresponding to quotients of atoms in $H_i$ by the edge-, the loop- and some choices of
half-quotients.}
\label{fig:reduction_expansion_diagram}
\end{figure}

We want to understand the \emph{expansion} of quotients, corresponding to the diagram in
Fig.~\ref{fig:reduction_expansion_diagram}. Suppose that $\Gamma_r$ is a semiregular subgroup of
$\Aut(G_r)$, defining the quotient $H_r = G_r / \Gamma_r$. The expansion constructs a series of
semiregular subgroups $\Gamma_r, \dots, \Gamma_0$ defining expanded quotients $H_i = G_i /
\Gamma_i$.

Unfortunately, the expansion is non-deterministic which means that the expanded quotients
$H_{r-1},\dots,H_0$ are not uniquely determined. Recall Lemma~\ref{lem:atom_covering_cases} and
Fig.~\ref{fig:projections_of_atoms}. The following characterization of all expanded
quotients is the main result of~\cite{fkkn16}: 

\begin{theorem}[\cite{fkkn16}, Theorem 1.2] \label{thm:quotient_expansion}
Let $G_{i+1}$ be a reduction of $G_i$.  Every quotient $H_i$ of $G_i$ can be constructed from some
quotient $H_{i+1}$ of $G_{i+1}$ by replacing each edge, loop and half-edge of $H_{i+1}$ by the
subgraph corresponding to the edge-, the loop-, or a half-quotient  of an atom of $G_i$,
respectively. 
\end{theorem}

\begin{figure}[t!]
\centering
\includegraphics{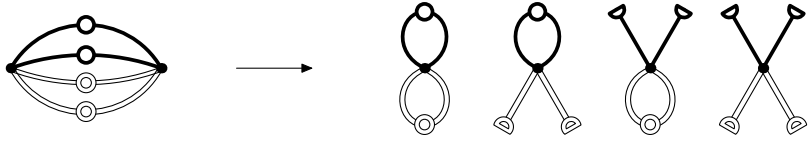}
\caption{An example of a dipole with four non-isomorphic half-quotiens.}
\label{fig:exponentially_many_quotients}
\end{figure}

By Lemma~\ref{lem:unique_quotients}, the edge and loop-quotients are uniquely determined. The
expansion is non-deterministic since there might be many non-isomorphic half-quotients, leading to
different graphs $H_i$. For instance, suppose that $H_{i+1}$ contains a half-edge corresponding to
the dipole from Fig.~\ref{fig:exponentially_many_quotients}. To construct $H_i$, we replace this
half-edge by one of the four possible half-quotients of this dipole.

\begin{corollary}[\cite{fkkn16}, Corollary 4.8] \label{cor:unique_expansion}
If $H_{i+1}$ contains no half-edge, then $H_i$ is uniquely determined. Thus, for an odd order of
$\Gamma_r$, the quotient $H_r$ uniquely determines $H_0$.
\end{corollary}

\heading{Half-quotients of Dipoles.}
Dipoles with colored edges may admit exponentially many non-isomorphic half-quotients; see
Fig.~\ref{fig:exponentially_many_quotients}. Therefore, if $H_{i+1}$ contains a half-edge
corresponding to a half-quotient of a dipole in $H_i$, the number of non-isomorphic expansions $H_i$
of $H_{i+1}$ can be exponential in the size difference of $H_i$ and $H_{i+1}$.

\begin{lemma}[\cite{fkkn16}, Lemma 4.9] \label{lem:dipole_quotients}
Let $A$ be a dipole with colored edges. Then the number of pairwise non-isomorphic half-quotients is
bounded by $2^{\lfloor {\be(A) / 2}\rfloor}$ and this bound is achieved. 
\end{lemma}

For the purpose of Section~\ref{sec:algorithm}, we describe the structure of all quotients of a
dipole. Each is constructed from an involutory semiregular automorphisms $\tau$ acting on $\int A$.
If $\bo A = \{u,v\}$, the half-quotient $A / \left<\tau\right>$ consists of a vertex $p(u)$ with several
loops and half-edges attached at $u$.  Since $\tau$ preserves the color classes and edge types, it
acts indepedently on each color class and type of edges.  

On each color class of the non-halvable edges of $A$, $\tau$ acts as a fixed-point free involution.
The undirected edges have to be paired by $\tau$ together. Each directed edge has to be paired with a
directed edge of the opposite direction and the same color.  In the quotient $A /
\left<\tau\right>$, we have no freedom: we get a (directed) loop for each such pair.

On each color class of halvable edges of $A$, $\tau$ acts as an arbitrary semiregular involution.
An edge $e$ fixed in $\tau$ is mapped into a half-edge of the given color in $A /
\left<\tau\right>$.  If $\tau$ maps $e$ to $e' \ne e$, then we get a loop in $A /
\left<\tau\right>$. The resulting half-quotient is therefore determined by  the numbers $h$ of fixed
edges and the numbers $\ell$ of two-cycles for each color class of halvable edges of size $m$ such
that $h+2\ell = m$.   

\heading{The Block Structure of Quotients.}
Last, we describe how the block structure changes during expansions.  A block atom $A$ of $G_i$ is
always projected by an edge-projection, so it corresponds to a block atom of $H_i$. Suppose that $A$
is a proper atom or a dipole with $\bo A = \{u,v\}$.  For an edge-projection, we get $p(u) \ne
p(v)$, and $p(A)$ is isomorphic to an atom in $H_i$.

For a loop- or half-projection, we get $p(u) = p(v)$ and $p(u)$ is an articulation of $H_i$.
If $A$ is a dipole, then $p(A)$ is a pendant star of half-edges and loops attached to $p(u)$. By
Lemma~\ref{lem:proper_atoms}, if $A$ is a proper atom, then $p(A)$ is either a path ending with a
half-edge and with attached single pendant edges (when $A^+$ is essentially a cycle), or a pendant
block with attached single pendant edges and half-edges (when $A^+$ is essentially 3-connected).
(The reason is that the fiber of an articulation in a 2-fold cover is a 2-cut.)

\begin{lemma}[\cite{fkkn16}, Lemma 4.10] \label{lem:block_structure}
The block structure of $H_{i+1}$ is preserved in $H_i$, possibly with some new subtrees of blocks
attached.
\end{lemma}

\section{Meta-algorithm} \label{sec:algorithm}

In this section, we establish the meta-algorithm from Theorem~\ref{thm:metaalgorithm}, solving
\rcover for $G$ belonging to $\calC$ satisfying (P1) to (P3) in time $\O^*(2^{\be(H)/2})$.

Let $k = |G|/|H|$, and we assume that $k \ge 2$. (If $k$ is not an integer, then clearly $G$ does
not cover $H$. If $k=1$, then it is equivalent to the graph isomorphism problem and we can test it
using Lemma~\ref{lem:testing_gi}.) The algorithm consists of the following major parts:
\begin{packed_enum}
\item \emph{Reduction Part:} We construct the reduction series for $G = G_0,\dots,G_r$ terminating
with the unique primitive graph $G_r$.  Throughout the reduction the central block is preserved,
otherwise according to Lemma~\ref{lem:preserved_center} there exists no semiregular automorphism of
$G$ and we output ``no''. According to (P1), the reduction preserves the class $\calC$, and also
every atom belongs to $\calC$.
\item \emph{Quotient Part:} Using (P2), we construct the list of all subgroups $\Gamma_r$ of
$\Aut(G_r)$ of the order $k$ acting semiregularly on $G_r$. The number of subgroups in the list is
polynomially large by (P2).
\item \emph{Expansion Part:} For each $\Gamma_r$ in the list, we compute $H_r = G_r / \Gamma_r$. We
say that a graph $H_r$ is \emph{expandable} if there exists a sequence of extensions repeatedly
applying Theorem~\ref{thm:quotient_expansion} which constructs $H_0$ isomorphic to $H$. We test the
expandability of $H_r$ using dynamic programming while using (P3).
\end{packed_enum}
It remains to explain details of Expansion Part, and prove the correctness of the algorithm.

\heading{Outline.} In Section~\ref{sec:algorithm_overview}, we give an overview of Expansion Part. In
Section~\ref{sec:catalog}, we describe a catalog which stores all atoms and their quotients
discovered during reductions. In Section~\ref{sec:reduction_with_lists}, we describe reductions with
lists, used in expandibility testing. Last, in Section~\ref{sec:proof_of_metaalgorithm}, we conclude
with a proof of Theorem~\ref{thm:metaalgorithm}.

\begin{figure}[t!]
\centering
\includegraphics{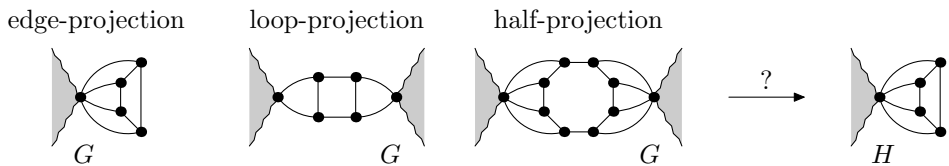}
\caption{For a pendant block of $H$, there are three possible preimages in $G$. It could be a block
atom mapped by the edge-projection, or a proper atom mapped by the loop-projection, or another
proper atom mapped by a half-projection (where the half-quotient is created by $180^\circ$ rotation
$\tau$).}
\label{fig:preimages_of_block}
\end{figure}

\subsection{Overview of Testing Expandability} \label{sec:algorithm_overview}

In this section, we explain how to test expandability of $H_r$.  We start by illustrating the
fundamental difficulty, for simplicity on pendant blocks. Suppose that $H$ has a pendant block as in
Fig.~\ref{fig:preimages_of_block}.  From the local information, there is no way to know whether
this block corresponds in $G$ to the edge-quotient of a block atom, or to the loop-quotients of some
proper atoms, or to half-quotients of some other proper atoms. It can easily happen that the
all these atoms appear in $G$. So without exploiting some additional information from $H$, there
is no way to know what is the preimage of this pendant block.

\begin{figure}[b!]
\centering
\includegraphics{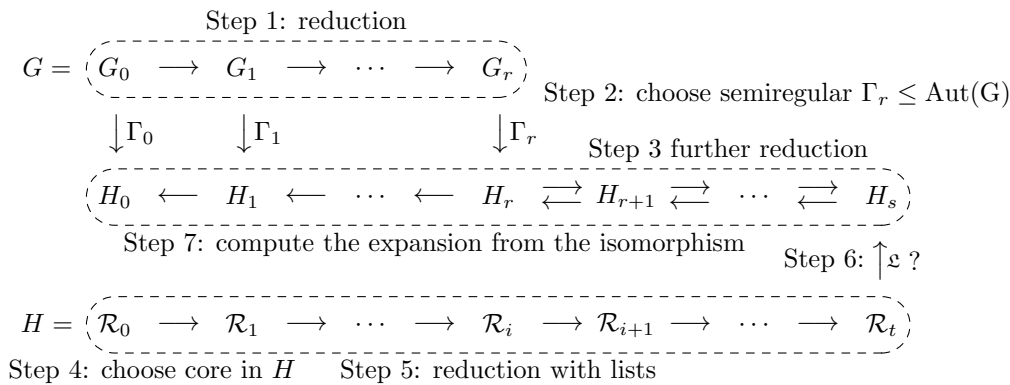}
\caption{The metaalgorithm proceeds in the following seven steps. We iterate over all possible
choices in Steps 2 and 4.}
\label{fig:metaalgorithm_diagram}
\end{figure}

In our approach, we revert the problem of expandability of $H_r$ by reducing $H$ towards $H_r$. But
since it is not clear which atoms of $H$ correspond to which parts of $G$, we do not decide it
during the reductions, instead we just remember lists of all possibilities.  The dynamic programming
deals with these lists and computes further lists for larger parts of $H$.
Figure~\ref{fig:metaalgorithm_diagram} illustrates the overview of our algorithm.

\begin{figure}[t!]
\centering
\includegraphics{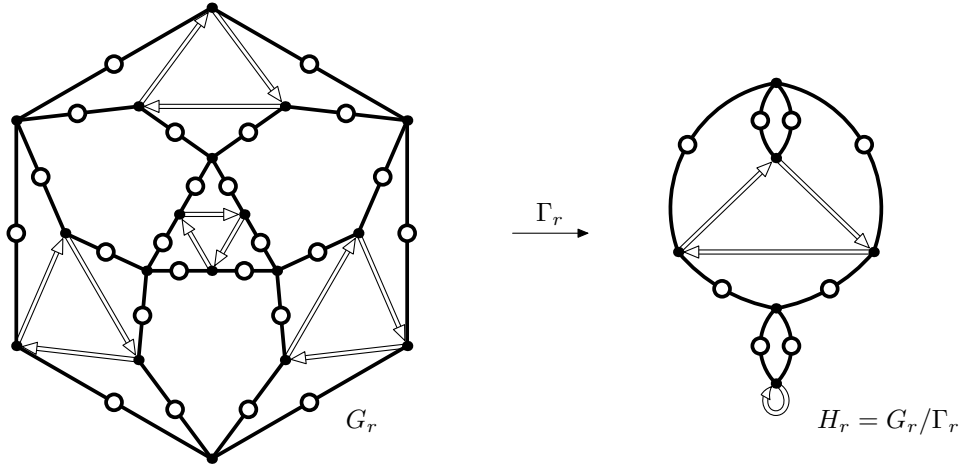}
\caption{A primitive graph $G_r$ which is 3-connected. Let $\Gamma_r$ be the semiregular subgroup of
$\Aut(G_r)$ generated by a 120$^\circ$ rotation. It defines the quotient $H_r = G_r / \Gamma_r$ which
is not primitive (contains articulations and 2-cuts).}
\label{fig:non_primitive_quotient}
\end{figure}

\heading{Reductions of Quotients and Cores.}
Notice that $H_r$ might not be primitive; see Fig.~\ref{fig:non_primitive_quotient} for an
example.  It would be difficult to match it to a reduction series in $H$, so in Step 3, we further
reduce $H_r$ to a primitive graph $H_s$.

\begin{figure}[b!]
\centering
\includegraphics{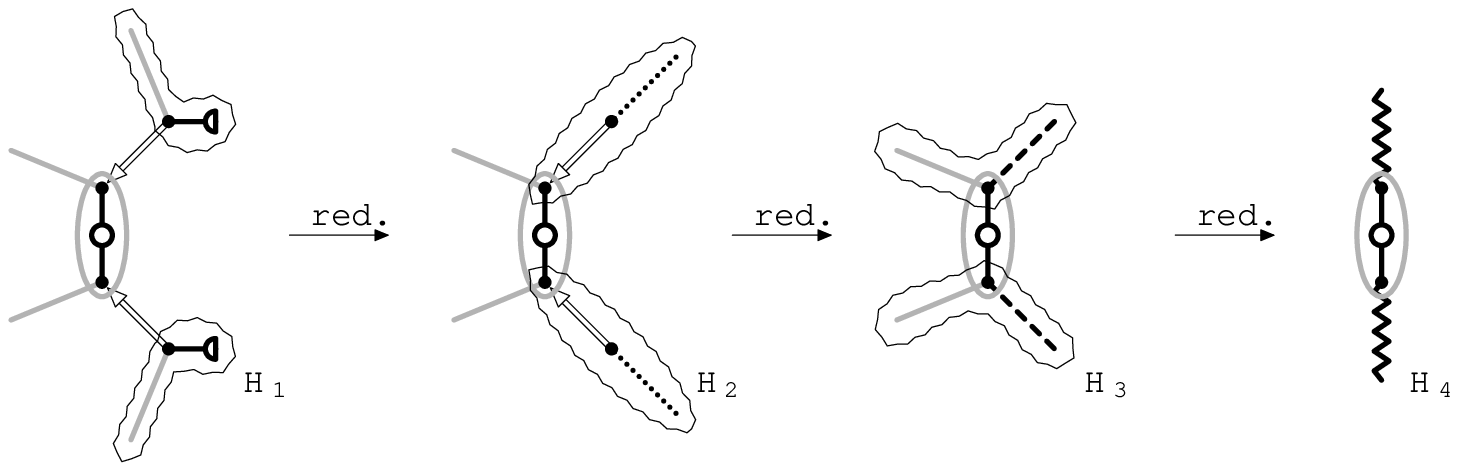}
\caption{The graph $H_1$ is one quotient of $G_1$ from Fig.~\ref{fig:example_of_reduction}. We
further reduce it to $H_3$ with respect to the core block depicted in gray. Notice that $H_1$ and
$H_2$ only contain block atoms.}
\label{fig:quotient_reduction}
\end{figure}

We define atoms in the quotient graphs similarly as in Section~\ref{sec:atoms} with only one
difference. We choose one arbitrary block/articulation called the \emph{core} in $H_r$; for
instance, we can choose the central block/articulation. The core plays the role of the central block
in the definition of parts and atoms. Also, in the definition we consider half-edges and loops as
pendant edges, so they do not form block atoms. We proceed with the reductions in $H_r$ further
till we obtain a primitive quotient graph $H_s$, for some $s \ge r$; see
Fig.~\ref{fig:quotient_reduction}.

Let $H_0,\dots,H_{s-1}$ be the graphs obtained by an expansion series of $H_s$ using
Theorem~\ref{thm:quotient_expansion}.

\begin{lemma} \label{lem:preserved_expandibility}
The graph $H_s$ is expandable to $H$, if and only if $H_r$ is expandable to $H$.
\end{lemma}

\begin{proof}
It follows from the fact that the graphs $H_{s-1},\dots,H_r$ are uniquelly determined, since no
half-edges are expanded till $H_r$.
\end{proof}

\begin{lemma} \label{lem:expanded_core}
If $H_s$ is expandable to $H$, then the core of $H_s$ is expanded to some block or articulation
of $H$.
\end{lemma}

\begin{proof}
The graph $H_s$ consists of the core together with some pendant edges, loops and half-edges.
By Lemma~\ref{lem:block_structure}, the core is preserved as an articulation/block in all graphs
$H_s,\dots,H_0$. The core can be only changed by replacing of its colored edges by edge-quotients.
Since $H_s$ is expandable to $H_0$, the expanded core is isomorphic to some block or articulation of
$H$.
\end{proof}

In Step 4, we test all possible positions of the core in $H$. (We have $\O(n)$ possibilities, so we
run the dynamic programming algorithm multiple times.) In what follows, we have the core fixed in
$H$ as well.

In Step 5, we apply on $H$ reductions with lists, described in
Section~\ref{sec:reduction_with_lists}. In Step 6, we test whether some choices from these lists are
compatible with the graph $H_s$.

\subsection{Catalog of Atoms} \label{sec:catalog}

During the reduction phase of the algorithm, we construct the following \emph{catalog of atoms}
forming a database of all discovered atoms. These atoms arise in three ways: atoms of
$G_0,\dots,G_{r-1}$, atoms in half-quotients of these atoms, and atoms in the reductions of the
quotients $H_r = G_r / \Gamma_r$. We are not very concerned with a specific implementation of the
algorithm, so the purpose of this catalog is to simplify description.

For each isomorphism class of atoms represented by an atom $A$, we store the following information
in the catalog:
\begin{packed_itemize}
\item The atom $A$.
\item The corresponding colored edge of a given type representing the atom in the reduction.
\item If $A$ is an atom of $G_0,\dots,G_{r-1}$, the unique edge- and loop-quotients of $A$ and
information about its half-quotients.
\end{packed_itemize}
For an overview of adding an atom $A$ into the catalog, see Algorithm~\ref{alg:catalog}.

\begin{algorithm}[b!]
\caption{The subroutine for adding an atom into the catalog} \label{alg:catalog}
\begin{algorithmic}[1]
\REQUIRE An atom $A$.
\ENSURE If $A$ is not contained in the catalog, then it is added. A colored
halvable/undirected/directed edge corresponding to $A$ is given.
\medskip
\IF {$A$ is a star block atom}
	\WHILE {$A$ contains a pendant edge $e'$ of a star block atom $S$}
		\STATE Replace $e'$ with the edges of $S$.
	\ENDWHILE
	\WHILE {$A$ contains a loop $e'$ of the loop-quotient of a dipole $D$}
		\STATE Replace $e'$ with the loops of the loop-quotient of $D$.
	\ENDWHILE
\ENDIF
\IF {$A$ is a dipole}
	\WHILE {$A$ contains an edge $e'$ corresponding to a dipole $D$}
		\STATE Replace $e'$ with the edges of $D$.
	\ENDWHILE
\ENDIF

\STATE We test whether $A$ is contained in the catalog using Lemma~\ref{lem:catalog_query}.
\IF {$A$ is contained in the catalog}
	\RETURN The corresponding colored edge representing $A$.
\ENDIF
\medskip
\STATE We determine the symmetry type of $A$ using Lemmas~\ref{lem:dipole_symmetry_type}
and~\ref{lem:proper_atom_symmetry_type}.
\STATE We assign an edge $e$ of a new color of the corresponding type to $A$.
\medskip
\IF {$A$ is an atom of $G_0,\dots,G_r$.}
	\STATE We compute the edge-quotient of $A$ and the loop-quotient of $A$ (if $A$ is not a block
	atom).	
	\IF {$A$ is a dipole consisting of exactly two halvable edges of the same color}
		\STATE We add the half-quotient of $A$ with one loop to the list of half-quotients.
	\ENDIF
	\medskip
	\IF {$A$ is a halvable proper atom}
		\STATE We compute all half-quotients $Q$ of $A$ by Lemma~\ref{lem:proper_atom_quotients}.
		\FOR {each half-quotient $Q$}
			\STATE Apply the reduction series on $Q$ with respect to the block containing $\bo Q$,
			constructing a primitive graph $Q'$.
			\STATE Add all detected atoms to the catalog and replace them by the corresponding
			colored edges.
			\STATE Add $Q'$ to the catalog, as a half-quotient of $A$. 
		\ENDFOR
	\ENDIF
\ENDIF
\medskip
\RETURN The assigned colored edge $e$ corresponding to $A$.
\end{algorithmic}
\end{algorithm}

\heading{Storing Star Block Atoms.}
Let $A$ be a star block atom. We store it in the catalog \emph{partially expanded} which works as
follows. By the definition, $A$ consists of a vertex with attached edges, loops and half-edges. If
some edge corresponds to a star block atom $S$, we replace it with the edges of $S$.  Similarly, if
some loop corresponds to the loop-quotient $Q$ of a dipole $D$, we replace it by the loops of $Q$.
We repeat this till all pendant edges of $A$ correspond to block atoms and all loops of $A$
correspond to loop-quotients of proper atoms. On the other hand, the half-edges of $A$ may
correspond to half-quotients of both proper atoms and dipoles.

\heading{Storing Dipoles.}
Let $A$ be a dipole in $G_0,\dots,G_{r-1}$. By Lemma~\ref{lem:dipole_quotients}, it can have
exponentially many non-isomorphic half-quotients. On the other hand, they are well described in
Section~\ref{sec:expansion}, so we can generate all of them from the dipole when needed.

We store this dipole $A$ in the catalog \emph{partially expanded} which works as follows. Almost all
edges of $A$ correspond to proper atoms, while at most one edge corresponds to a dipole $D$. (At most
one since from the definition, a dipole $D$ with $\bo D = \{u,v\}$ consists of all edges between $u$
and $v$.) If one edge corresponds to $D$, we replace it in $A$ with the edges of the dipole $D$. And
if one of these edges of $D$ again correspond to some dipole $D'$, we proceed further with the
expansion.

Notice that by the definition of the reduction, all colored edges of $D$ have different colors than
the edges of $A$. Therefore the half-quotients of the original dipole $A$ are exactly the same as
the half-quotients of the partially expanded dipole $A$. The reason for this expansion is that every
half-quotient of the partially expanded dipole $A$ consists of loops and half-edges attached to one
vertex, where each loop and each half-edge is expanded into one block (with attached single pendant
edges, half-edges and loops).

Further, if a halvable dipole $A$ consists of exactly two edges of the same color, we compute its
half-quotient consisting of just the single loop attached, and we add this quotient to the catalog.
The reason is that this quotient behaves exactly as the loop-quotient of some proper atom.

\begin{figure}[t!]
\centering
\includegraphics{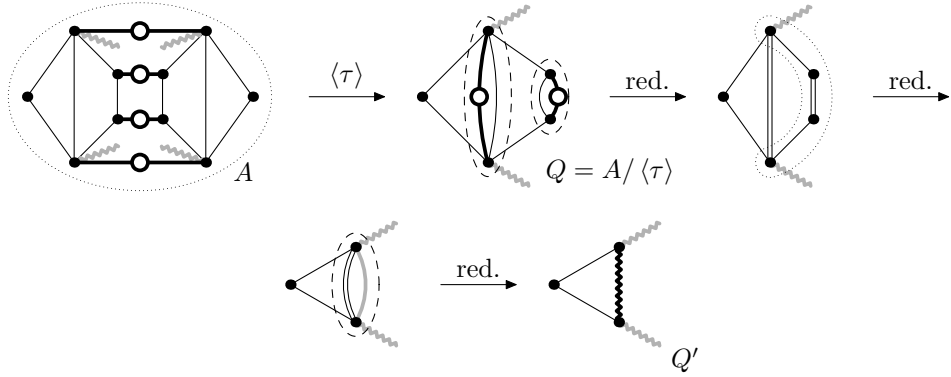}
\caption{A proper atom $A$ with a half-quotient $Q$ generated by 180$^\circ$ rotation $\tau$.
A reduction series is applied on $Q$ which adds further atoms to the catalog and the primitive graph
$Q'$.}
\label{fig:catalog_quotient_reduction}
\end{figure}

\heading{Storing Proper Atoms.}
If $A$ is not a dipole, we compute the list of all its pairwise non-isomorphic half-quotients, and
store them in the catalog in the following way. A half-quotient $Q$ of $A$ might not be primitive.
Therefore, we apply a reduction series on $Q$, and add all atoms discovered by the reduction to the
catalog.  (We do not compute their half-quotients. They are never realized unless these atoms are
directly found in $G$ as well.) When the reduction series finishes, this half-quotient is reduced to
a primitive graph $Q'$. Naturally, the block containing $\bo Q$, being a single vertex of the
half-quotient, behaves like the central block in the definition of atoms, i.e., it is never reduced.
The reduced half-quotient $Q'$ is either essentially 3-connected, a cycle with attached single
pendant edges, or $K_2$ with a single pendant edge or half-edge attached. See
Fig.~\ref{fig:catalog_quotient_reduction} for an example.

\heading{Total Size of Catalog.} Next, we prove that the catalog is not too large.

\begin{lemma} \label{lem:catalog_size}
Assuming (P2), the catalog contains polynomially many atoms and half-quotients.
\end{lemma}

\begin{proof}
First we deal with the number of atoms in $G_0,\dots,G_r$. Notice that by replacing an interior of
an atom, the total number of vertices and edges is decreased; the interiors of atoms in each $G_i$
contain at least two vertices and edges in total and are pairwise disjoint (see
Lemma~\ref{lem:nonintersecting_atoms}). Thus we add a linear number of atoms of $G_0,\dots,G_r$ to
the catalog, of total linear size.

By (P2), there are polynomially many possible quotients $H_r$, in each we encounter linearly many atoms
when reducing to $H_s$. So we add polynomially many atoms to the catalog.

By (P2), each proper atom $A$ has polynomially many half-quotients, for different semiregular
involutions of $\Aut(A)$. So, we have in total polynomially many half-quotients, each containing at
most linearly many atoms in its reduction series.  And by Lemma~\ref{lem:unique_quotients} we have
the unique edge- and loop-quotient. So again, the total number of atoms and quotients added to the
catalog is polynomial.
\end{proof}

\heading{Catalog Queries.}
Throughout the algorithm, we repeatedly ask \emph{queries} whether some atom or some of its
quotients is contained in the catalog, and if so, we retrieve the corresponding colored
edge/loop/half-edge.

\begin{lemma} \label{lem:catalog_query}
Assuming (P3$\,{}^*\!$), each catalog query can be answered in polynomial time.
\end{lemma}

\begin{proof}
By Lemma~\ref{lem:catalog_size}, we need to test graph isomorphism for the input atom/quotient and
polynomially many atoms/quotients in the catalog. If the input is an atom of an edge-quotient, we use
Lemma~\ref{lem:testing_gi_atoms}. If it is a loop- or a half-quotient, then it is a primitive graph
and we use Lemma~\ref{lem:testing_gi_primitive_graphs}.
\end{proof}

\subsection{Reductions with Lists} \label{sec:reduction_with_lists}

In this section, we describe Steps 5 and 6 of the diagram in Fig.~\ref{fig:metaalgorithm_diagram}.
By Lemma~\ref{lem:preserved_expandibility}, we need to test whether $H_s$ is expandible to $H$.  We
approach this in the opposite way, by applying a reduction series on $H$ with respect to the core
defining $\calH_0,\dots,\calH_t$. As already discussed in Section~\ref{sec:algorithm_overview}, we
do not know which parts of $G$ project to different parts of $H$. Therefore each $\calH_i$ is a set
of graphs, and $\calH_t$ is a set of primitive graphs. We then determine expandability of $H_s$
by testing whether $H_s \in \calH_t$.

Since each set $\calH_i$ can contain a huge number of graphs, we represent it implicitly in the
following manner. Each $\calH_i$ is represented by one graph $\calR_i$ with some colored edges and
with so-called pendant elements attached to some vertices. 

\heading{Pendant Elements with Lists.}
A pendant element $x$ in $\calR_i$ corresponds to a block atom in $\calR_j$ for some $j < i$, which
is reduced in $\calR_{j+1}$. When pendant elements are fully expanded, they correspond to block part
of $H$ with pairwise disjoint interiors. We use the name pendant element since it may represent a
pendant edge of some color, several loops of some other colors, and several half-edges of some other
colors. 

\begin{figure}[t!]
\centering
\includegraphics{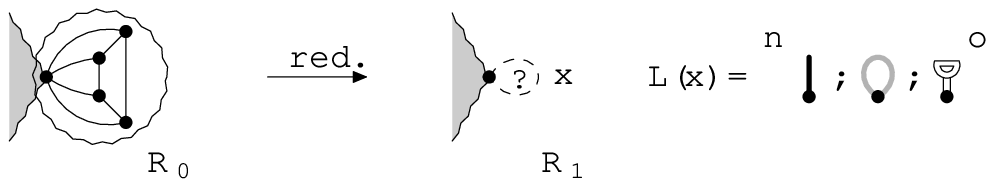}
\caption{Let $x$ be the pendant element corresponding to the pendant block of $H$ depicted in
Fig.~\ref{fig:preimages_of_block}. Then $\frakL(x)$ contains three different members if all three
atoms depicted in Fig.~\ref{fig:preimages_of_block} are contained in the catalog.}
\label{fig:list_example}
\end{figure}

Each pendant element $x$ is equipped with a \emph{list} $\frakL(x)$ whose \emph{members} are
possible realizations of the corresponding block atom by the quotients from the catalog. Each graph
of $\calH_i$ is created for $\calR_i$ by replacing the pendant elements by some choices of edges,
loops and half-edges from their lists. The list $\frakL(x)$ of a pendant element $x$ contains an
edge/loop/half-edge if and only if it is possible to expand this edge/loop/half-edge to the graph
isomorphic to the block part corresponding to $x$. For an example, see
Fig.~\ref{fig:list_example}. According to Lemma~\ref{lem:catalog_size}, we have polynomially many
atoms, and so the size of each list is polynomial in size. 

\begin{lemma} \label{lem:list_properties}
Each list $\frakL(x)$ contains at most one edge. Further, if two lists share an edge or a loop,
their pendant elements correspond to isomorphic block parts in $H$.
\end{lemma}

\begin{proof}
Two atoms have the isomorphic edge-quotients if and only if they are isomorphic. Therefore each
list $\frakL(x)$ contains at most one edge.

If a pendant element $x$ is fully expanded, it corresponds to one block part of $H$. Suppose that an
edge- or a loop-quotient belongs to $\frakL(x)$. If it is fully expanded, then it has to be isomorphic to
this block part. But according to Lemma~\ref{lem:unique_quotients}, the expansions of edge- and
loop-quotients are deterministic since half-edges are never encountered. Therefore the corresponding
block part in $H$ is uniquely determined.
\end{proof}

\begin{figure}[b!]
\centering
\includegraphics{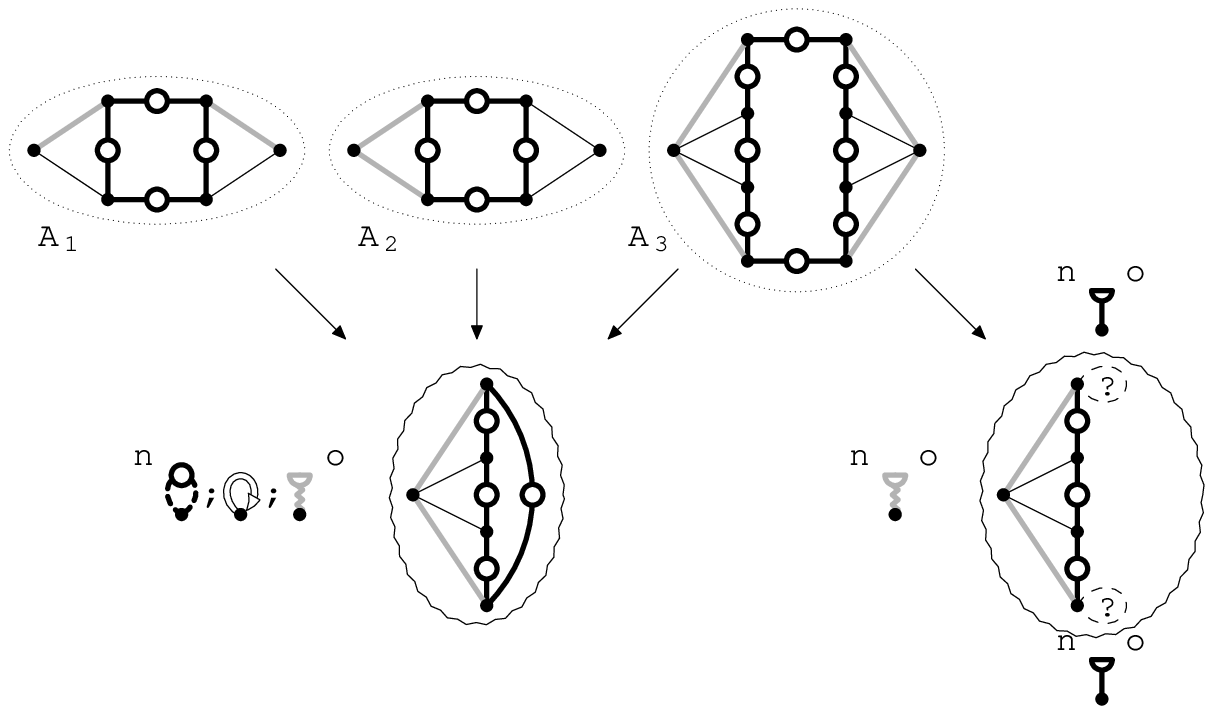}
\caption{Two block atoms corresponding to pendant elements with depicted lists. On the left, the list 
has the loops corresponding to $A_1$ and $A_2$ and the half-edge corresponding to $A_3$.
On the right, the list only contains the half-edge of $A_3$.}
\label{fig:lists_sharing}
\end{figure}

\begin{figure}[t!]
\centering
\includegraphics{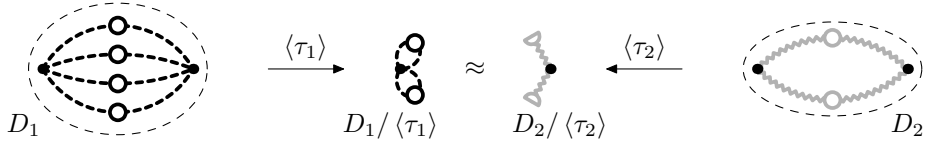}
\caption{An example of two dipoles $D_1$ and $D_2$ having isomorphic half-quotients. Consider the
atoms $A_1$ and $A_3$ from Fig.~\ref{fig:lists_sharing}, for which the loop-quotient of $A_1$ is
isomorphic to a half-quotient of $A_3$. Let $D_1$ consist of four edges corresponding to $A_1$ and
let $D_2$ consist of two edges corresponding to $A_3$. Then the half-quotient $D_1 /
\left<\tau_1\right>$ can be expanded to a graph isomorphic to an expansion of the half-quotient $D_2/\left<\tau_2\right>$.}
\label{fig:list_two_halfedges}
\end{figure}

One list may contain several loops, for which identifying of the vertices of the boundaries
constructs identical graphs; see Fig.~\ref{fig:lists_sharing}.  Similarly, a list may contain
several half-edges; see Fig.~\ref{fig:list_two_halfedges}. Because of the second part of
Lemma~\ref{lem:list_properties}, the loops pose no problem. On the other hand, one half-edge may be
contained in lists of several different pendant elements which are expanded to non-isomorphic
subgraphs in $H$; see Fig.~\ref{fig:lists_sharing}.  This creates the main difficulty for our
algorithm, leading to the bottleneck in form of a slow subroutine requiring time $\O^*(2^{\be(H)/2})$. 

\heading{Reductions with Lists.}
We want to compute the reduction series with lists $H = \calR_0, \calR_1, \dots, \calR_t$ ending
with a primitive graph $\calR_t$ with attached pendant elements with computed lists.  We construct
$\calR_0$ by replacing all pendant edges and loops by pendant elements with singleton lists.

Suppose that we know $\calR_i$, and we want to apply one step of the reduction and compute
$\calR_{i+1}$. We find all atoms in $\calR_i$. We define atoms with respect to the chosen
core in $H$, and we work with pendant elements as with pendant edges. Further, we consider only star
block atoms consisting only of an articulation with all its descendants attached in form of reduced
pendant elements. This means that we postpone reduction of star block atoms till all their
descendants are reduced first. (The same modification could be applied in all reductions as well,
but it is important here.)

To construct $\calR_{i+1}$ from $\calR_i$, we proceed with the following:
\begin{packed_itemize}
\item We replace dipoles and proper atoms by edges of the corresponding colors from the catalog.  A
proper atom might have pendant elements attached to its interior, but these pendant elements are always
realized by edges corresponding to the edge-quotients of some block atoms. Therefore, we can replace
the pendant elements with lists by the unique edges from these lists, and if some list contains no
edge, we stop the reduction. We run a catalog query and if the dipole or proper atom is not
contained in the catalog, we halt the reduction procedure. 
\item We replace block atoms by pendant elements with constructed lists.  If some list is empty, we
again halt the reduction.
\end{packed_itemize}
It remains to describe the construction of the lists for the created pendant elements.

\heading{Computing Lists.} Let $A$ be a block atom in $\calR_i$, replaced by a pendant element $x$
in $\calR_{i+1}$, and we want to compute $\frakL(x)$. We compute $\frakL(x)$ from the lists of the
pendant elements attached to $A$. Suppose that $A$ has pendant elements $y_1,\dots,y_p$ attached.  For
each member of $\frakL(x)$, we remember which members of $\frakL(y_1),\dots,\frakL(y_p)$ have to be
chosen for its expansion.

\begin{lemma} \label{lem:lists_for_nonstar_atoms}
Let $A$ be a non-star block atom in $\calR_i$. Assuming (P3), we can compute the list $\frakL(x)$ of
the pendant element $x$ corresponding to $A$ in polynomial time.
\end{lemma}

\begin{figure}[t!]
\centering
\includegraphics{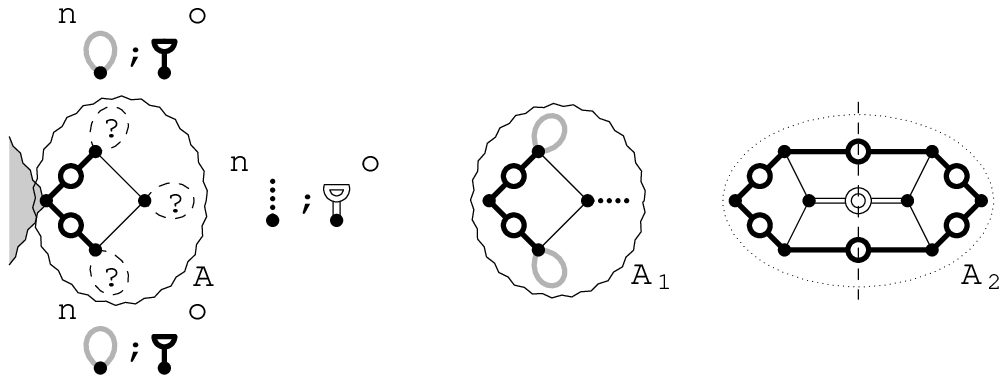}
\caption{On the left, a non-star block atom $A$ in $\calR_i$ with depicted lists of its pendant
elements.  On the right, two possible atoms from the catalog having a quotient for which there
exists a list-compatible isomorphism from $A$. So the list of the pendant element replacing $A$ in
$\calR_{i+1}$ contains the pendant edge corresponding to the edge-quotient of the block atom $A_1$
and the half-edge corresponding to a half-quotient of the proper atom $A_2$. }
\label{fig:lists_for_nonstar_atoms}
\end{figure}

\begin{proof}
We iterate over quotients in the catalog which are $K_2$ with a single pendant edge, essentially
cycles, or essentially 3-connected graphs by Lemma~\ref{lem:non_star_block_atoms}.  Let $Q$ be such a
quotient.  For $u \in \bo A$, we put $\frakL(u) = \bo Q$.  For each single pendant element $y$ of
$A$ attached at $u$, we construct $\frakL(u)$ consisting of all vertices $v \in \bV(Q)$ such that
the pendent edge/loop/half-edge attached at $v$ belongs to $\frakL(y)$. We remove all pendant
elements attached at $A$ and all pendant edges/loops/half-edges attached at $Q$. 

By definition of pendant elements, it is possible to expand $Q$ to the block part corresponding to
$A$ if and only if there exists a list-compatible isomorphism $A \listiso Q$. If $Q$ is $K_2$ or a
cycle, we test it trivially. If $Q$ is 3-connected, we test it using (P3). If $A \listiso Q$, we add
the pendant edge/loop/half-edge representing this quotient to the list $\frakL(x)$, and we remember
the constructed isomorphism $A \listiso Q$.  See Fig.~\ref{fig:lists_for_nonstar_atoms} for an
example. 
\end{proof}

On the other hand, if $A$ is a star block atom, we compute its list by a slow subroutine. If this
slow subroutine can be avoided and the list for $A$ can be computed in polynomial time, the entire
meta-algorithm of Theorem~\ref{thm:metaalgorithm} runs in polynomial time.

\begin{lemma} \label{lem:lists_for_star_atoms}
Let $A$ be a star block atom in $\calR_i$. We can compute the list $\frakL(x)$ of the pendant element $x$
corresponding to $A$ in time $\O^*(2^{\be(H)/2})$.
\end{lemma}

\begin{proof}
Each star block atom of $\calR_i$ corresponds either to the edge-quotient of a star block atom, or
to the loop- or a half-quotient of a dipole. Lemma~\ref{lem:dipole_quotients} states that a dipole
can have exponentially many pairwise non-isomorphic half-quotients, we iterate over all 
of them which gives $2^{\be(H)/2}$ part in the complexity bound. Since we postpone reduction of star
block atoms, all pendant elements of $A$ necessarily correspond to non-star block atoms in
some $\calR_j$, for $j < i$, so each pendant element corresponds in $H$ to one subtree of blocks attached
at the vertex of $A$.

\emph{Case 1: Dipoles.}
We iterate over all partially expanded dipoles in the catalog and try to add them to the list
$\frakL(x)$. Let $D$ be a partially expanded dipole, recall that all edges of $D$ correspond to
proper atoms.

We test whether the lists of the pendant elements attached to the star block atom $A$ are compatible
with the loop-quotient of $D$. Each loop of this loop-quotient corresponds to the loop-quotient of
some proper atom which is either a cycle with attached single pendant edges, or essentially
3-connected by Lemma~\ref{lem:proper_atoms}. Therefore, it corresponds to exactly one pendant
element in $A$.  By Lemma~\ref{lem:list_properties}, each loop belongs only to lists of pendant
elements of type. Therefore, we just need to compare the number of loops in each color class with
the number of lists containing this colored loop. If these numbers match, we add the loop
representing the loop-quotient of $D$ to $\frakL(x)$.

Then we iterate over all half-quotients of $D$. By Lemma~\ref{lem:dipole_quotients}, let $Q$ be one
of its at most $2^{\be(H)/2}$ possible quotients. Recall from Section~\ref{sec:structural_results}
that an edge of $D$ projects either to a half-edge, or together with another edge of $D$ of the same
color and type to one loop. So each $Q$ consists of loops and half-edges attached to a vertex.
Since all edges of $D$ correspond to proper atoms, each loop and each half-edge has to be matched to
one pendant element of $A$.

\begin{figure}[b!]
\centering
\includegraphics{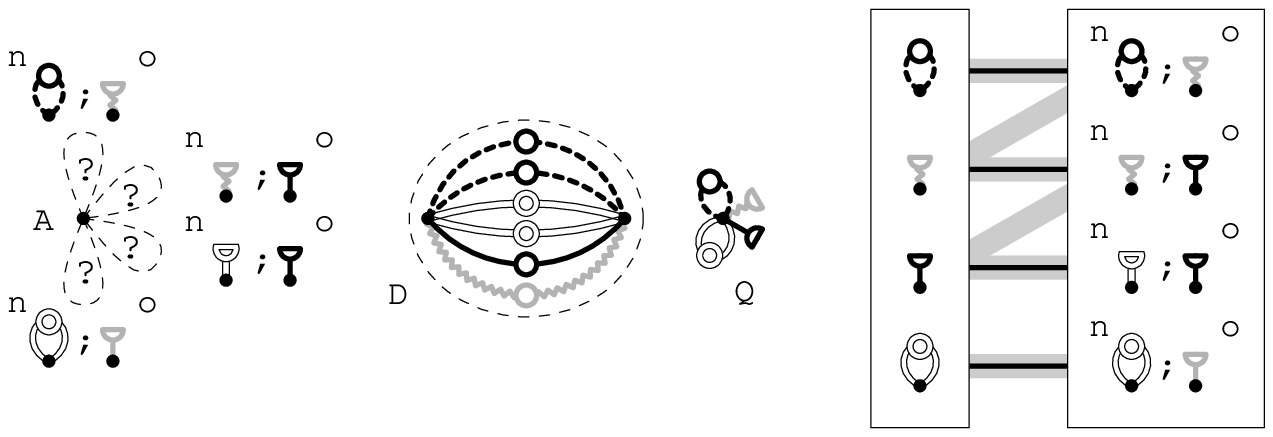}
\caption{The half-edge corresponding to a half-quotient of the dipole $D$ belongs to the list
$\frakL(x)$ of a pendant element $x$ replacing $A$ because there exists a perfect matching between
the loops and half-edges of the half-quotient $Q$ of $D$ and the lists of pendant elements of $A$.}
\label{fig:lists_for_star_atoms}
\end{figure}

Therefore, we test existence of a perfect matching in the following bipartite graph: One part is
formed by the loops and the half-edges of $Q$, and the other part is formed by the pendant elements
of $A$. A loop/half-edge is adjacent to a pendant element, if and only if the corresponding list
contains this loop/half-edge. Each perfect matching defines one assignment of the loops and
half-edges of $Q$ to the pendant elements of $A$. See Fig.~\ref{fig:lists_for_star_atoms} for an
example. We add the half-edge corresponding to a half-quotient of $D$ to $\frakL(x)$ if and only if
there exists a perfect matching for at least one half-quotient $Q$ of $D$.

\emph{Case 2: Star Block Atoms.}
We iterate over all partially expanded star block atoms of the catalog, let $S$ be one of them.  The
star block atom $S$ consists of one vertex with attached pendant edges (corresponding to non-star
block atoms), loops (corresponding to the loop-quotients of proper atoms) and half-edges. Some of
these half-edges correspond to dipoles, and some to proper atoms. Let $h_1,\dots,h_d$ be the
half-edges corresponding to partially expanded dipoles $D_1,\dots,D_d$ from the catalog. We
construct all expanded edge-quotients $Q$ of $S$ by replacing $h_1,\dots,h_d$ by all possible
choices of half-quotients $Q_1,\dots,Q_d$ of $D_1,\dots,D_d$.  In total, we have at most
$2^{\be(H)/2}$ different expanded edge-quotients $Q$ of $S$.

All pendant edges of $Q$ correspond to non-star block atoms, and all loops and half-edges correspond to
loop- and half-quotients of proper atoms. Therefore, every edge, loop and half-edge attached in $Q$
has to be matched to one pendant element of $A$.  Similarly as above, for each expanded
edge-quotient $Q$, we test whether there exists a perfect matching between edges, loops and
half-edges of $Q$ and the lists of pendant elements of $A$. We add the edge representing the
edge-quotient of the star block atom $S$ to $\frakL(x)$, if and only if there exists a perfect
matching for some expanded edge-quotient $Q$ of $S$.

The procedure computes the list $\frakL(x)$ correctly since we test all possible quotients from the
catalog, and for each quotient we test all possibilities how it could be matched to $A$.  For each
quotient $Q$, the running time is clearly polynomial, and we have $\O^*(2^{\be(H)/2})$ quotients.
\end{proof}

Algorithm~\ref{alg:computing_list} gives the pseudocode for computation of the list $\frakL(x)$ of a
pendant element $x$ replacing an atom $A$. If the returned list is empty, we halt the reduction;
either $H_s$ is not expandable to $H$, or we have chosen a wrong core in $H$.

\begin{algorithm}[t!]
\caption{The subroutine for computing lists of pendant elements} \label{alg:computing_list}
\begin{algorithmic}[1]
\REQUIRE A block atom $A$ of $\calR_i$.
\ENSURE The list $\frakL(x)$ of the pendant element $x$ replacing $A$ in $\calR_{i+1}$.
\medskip
\STATE Initiate the empty list $\frakL(x)$.
\IF {$A$ is a non-star block atom}
	\STATE Iterate over all quotients from the catalog.
	\FOR{each quotient $Q$ from the catalog}
		\STATE For each pendant element $x$ of $A$ attached at $u$, set $\frakL(u)$ to all vertices
		$\bV(Q)$ having an attached pendant edge/loop/half-edges belonging to $\frakL(x)$, and remove
		$x$.
		\STATE Remove all pendant edges/loops/half-edges in $Q$, and test $A \listiso Q$ (trivially or using (P3)).
		\STATE If some list-compatible isomorphism exists, we add the edge/loop/half-edge of $Q$ to
		$\frakL(x)$ together with this list-compatible isomorphism.
	\ENDFOR
\ENDIF
\medskip
\IF {$A$ is a star block atom}
	\STATE Iterate over all partially expanded dipoles $D$ and star block atoms $S$ in the catalog.
	\FOR{each partially expanded dipole $D$}
		\STATE Test whether the loop-quotient of $D$ matches the lists; if yes, then add the loop
		representing $D$ to $\frakL(x)$.
		\STATE Iterate over all half-quotients $Q$ of $D$.
		\FOR{each half-quotient $Q$}
			\STATE Test existence of a perfect matching between the loops and half-edges of $Q$ and
			the lists of the pendant elements of $A$.
			\STATE If a perfect matching exists, add the half-edge of $D$ to $\frakL(x)$ together
			with this half-quotient $Q$ and this matching, and proceed with the next dipole.
		\ENDFOR
	\ENDFOR
	\FOR{each partially expanded star block atom $S$}
		\STATE Compute all expanded edge-quotients $Q$ of $S$ by replacing the half-edges $h_1,\dots,h_d$
		corresponding to the dipoles by all possible combinations of their half-quotients $Q_1,\dots,Q_d$.
		\FOR{each expanded edge-quotient $Q$}
			\STATE Test existence of a perfect matching between the edges, loops and half-edges of $Q$
			and the lists of the pendant elements of $A$.
			\STATE If a perfect matching exists, then add the edge of $S$ to $\frakL(x)$ with this
			expanded edge-quotient $Q$ and this matching, and proceed with the next star block atom.
		\ENDFOR
	\ENDFOR
\ENDIF
\medskip
\RETURN The constructed list $\frakL(x)$.
\end{algorithmic}
\end{algorithm}

\heading{Testing Expandibility.}
The reduction with lists ends with a primitive graph $\calR_t$ with lists. For one particular choice
of a core, $\calR_t$ represents the set of graphs $\calH_t$ to which $H$ can be reduced.

In the following, we denote by $\calR_t \listiso H_s$ existence of a \emph{list-compatible
isomorphism} which preserves colors and orientations of edges and maps pendant elements $x$ of
$\calR_t$ into pendant edges, loops and half-edges of $H_s$ such that $\pi(x) \in \frakL(x)$. (It
corresponds to a list-compatible isomorphism defined in Section~\ref{sec:introduction} when pendant
elements/edges/loops/half-edges are removed, as described in the proof of
Lemma~\ref{lem:lists_for_nonstar_atoms}.)

\begin{lemma} \label{lem:expandability_correctness}
The graph $H_s$ is expandable to $H_0$ which is isomorphic to $H$ if and only if $\calR_t \listiso
H_s$ for some choice of the core in $H$.
\end{lemma}

\begin{proof}
Suppose that $\calR_t \listiso H_s$ for some choice of the core.  By the definition, every
pendant element $x$ of $\calR_t$ can be replaced by any member of $\frakL(x)$ which can be fully
expanded to the block part in $H$ corresponding to $x$.  The list-compatible isomorphism chooses for
pendant elements of $\calR_t$ realization by edges, loops and half-edges which is compatible with
the computed quotient of $H_s$. 

In more detail, we first expand edges in $H_s,\dots,H_{r+1}$ by the unique edge-quotients to reach
$H_r$, this has to be compatible with the sequence of replacements defined by $\calR_t \listiso
H_s$. Then we do replacements in the manner of Theorem~\ref{thm:quotient_expansion}, and construct
the expansions $H_{r-1},\dots,H_0$. Since we expand according to the list-compatible isomorphism $
\calR_t \listiso H_s$, the constructed graph $H_0$ is isomorphic to $H$.

On the other hand, suppose that $H_s$ is expandable to $H_0$ which is isomorphic to $H$. Then
according to Lemma~\ref{lem:block_structure}, the core of $H_s$ is preserved in $H$, so it has to
correspond to some block or to some articulation of $H$, which we choose as the core of $H$.  Since
there exists a sequence of replacements from $H_s$ which constructs $H_0$, this sequence of
replacements is possible in $\calR_t$. Thus $\calR_t \listiso H_s$.
\end{proof}

\begin{lemma} \label{lem:testing_expandibility}
Assuming (P3), we can test whether $H_s$ is expandable to $H_0$ which is isomorphic to $H$ in time
$\O^*(2^{\be(H)/2})$.
\end{lemma}

\begin{proof}
We iterate over all choices of the core in $H$. For each, we compute the reduction series with lists
$H = \calR_0, \dots, \calR_t$, using Algorithm~\ref{alg:computing_list} and
Lemmas~\ref{lem:lists_for_nonstar_atoms} and~\ref{lem:lists_for_star_atoms}.  We modify both graphs
$\calR_t$ and $H_s$ similarly as in the proof of Lemma~\ref{lem:lists_for_nonstar_atoms}.  For each
pendant element $x$ of $\calR_t$ attached at $u$, we put $\frakL(u)$ be the set of all vertices of
$\bV(H_s)$ having an attached pendant edge/loop/half-edge which belongs to $\frakL(x)$, and we
remove $x$. We remove all pendant edges/loops/half-edges of $H_s$.

Then we test whether $\calR_t \listiso H_s$.  It is trivial to deal with the cases when $H_s$ or
$\calR_t$ are cycles, $K_2$ or $K_1$. Otherwise by Lemma~\ref{lem:primitive_graphs}, both $H_s$ and
$\calR_t$ are 3-connected graphs, and we test $\calR_t \listiso H_s$ using (P3). By
Lemma~\ref{lem:expandability_correctness}, this subroutine is correct and runs in time
$\O^*(2^{\be(H)/2})$.
\end{proof}

\subsection{Proof of The Main Theorem} \label{sec:proof_of_metaalgorithm}

Now, we are ready to establish the main algorithmic result of the paper; see
Algorithm~\ref{alg:metaalgorithm} for the pseudocode. Assuming that a class $\calC$ satisfies (P1)
to (P3), we show that \rcover\ can be solved for $\calC$-inputs $G$ in time $\O^*(2^{\be(H)/2})$:

\begin{proof}[Theorem~\ref{thm:metaalgorithm}]
We recall the main steps of the algorithm and discuss their time complexity.  The reduction series
$G_0,\dots,G_r$ can be computed in polynomial time, by Lemmas~\ref{lem:reduction_series}
and~\ref{lem:catalog_query}. We reach in $G_r$ one of primitive graphs characterized in
Lemma~\ref{lem:primitive_graphs}.  If $G_r$ is essentially 3-connected, the property (P2) ensures
that there are polynomially many semiregular subgroups $\Gamma_r$ of $\Aut(G_r)$ which can be
computed in polynomial time. If $G_r$ is $K_2$ with attached single pendant edges or essentially a
cycle, it is true as well.

For each of these subgroups $\Gamma_r$, we compute the quotient $H_r = G_r / \Gamma_r$.  Then we
compute the reduction series $H_r,\dots,H_s$, again in polynomial time using
Lemma~\ref{lem:catalog_query}. Using Lemma~\ref{lem:testing_expandibility}, we test in time
$\O^*(2^{\be(H)/2})$ whether $H_s$ is expandable to $H_0$ which is isomorphic to $H$. We output
``yes'' if and only if $H_r = G_r / \Gamma_r$ is expandable to $H_0$ isomorphic to $H$ for at least
one the subgroups $\Gamma_r$.

To certify the ``yes'' outputs, we construct the semiregular subgroup $\Gamma \le \Aut(G)$ such that $G
/ \Gamma \cong H$ as follows. If $\calR_t \listiso H_s$ for some choice of the core in $H$, this
list-compatible isomorphism describes how to expand $H_s$ to $H_0$ which is isomorphic to $H$ using
Theorem~\ref{thm:quotient_expansion}. This expansion replaces edges, loops and half-edges with
edge-quotients, loop-quotients and some choices of half-quotients. The expansion towards $H_r$ is
deterministic since no half-edges are replaced.

We expand $H_r,\dots,H_0$, together with constructing group extensions $\Gamma_{r-1},\dots,\Gamma_0$
of $\Gamma_r$, where $\Gamma_i$ is a semiregular subgroup of $\Aut(G_i)$. When $G_{i+1}$ is expanded
to $G_i$, we replace some edges with interiors of some atoms. In the common parts, we define the
actions of $\Gamma_i$ and $\Gamma_{i+1}$ the same.  It remains to define the action of $\Gamma_i$ on
the interiors of these atoms in such a way that $G_i / \Gamma_i = H_i$. When some orbit of atoms is projected to
the edge- or loop-quotients, then we isomorphically swap their interiors in $\Gamma_i$ the same as
the corresponding edges are swapped in $\Gamma_{i+1}$.  For an orbit which is projected to
half-quotients, we further compose some automorphisms of $\Gamma_i$ with the half-quotient defining
semiregular involution $\tau$ on their interior (when the corresponding edges are flipped in
$\Gamma_i$). For more details, see~\cite{fkkn16}, proofs of Lemma~4.7 and Theorem~1.3 therein.

It remains to argue correctness of the algorithm. First suppose that the algorithm succeeds.  We
construct a semiregular subgroup $\Gamma$ of $\Aut(G)$. By
Lemma~\ref{lem:expandability_correctness}, some $H_s$ is expandible to $H_0$ which is isomorphic to
$H$. By Theorem~\ref{thm:quotient_expansion}, we get that $G / \Gamma \cong H$ which proves that $G$
regularly covers $H$. On the other hand, suppose that there exists a semiregular $\Gamma$ such that
$H \cong G / \Gamma$. Then $\Gamma$ corresponds to the unique semiregular subgroup $\Gamma_r$ on
$G_r$ which is one of the semiregular subgroups tested by the algorithm. Therefore $H_r$ has to be
expandable to $H_0$ isomorphic to $H$, and we detect this correctly according to
Lemma~\ref{lem:expandability_correctness}.
\end{proof}

\begin{algorithm}[t!]
\caption{The meta-algorithm for regular covers -- $\rcover$} \label{alg:metaalgorithm}
\begin{algorithmic}[1]
\REQUIRE A graph $G$ of $\calC$ satisfying (P1), (P2) and (P3), and a graph $H$.
\ENSURE A semiregular subgroup $\Gamma \le \Aut(G)$ such that $G / \Gamma \cong H$ if it exists.
\medskip
\STATE Compute the reduction series $G_0,\dots,G_r$ ending with the primitive graph $G_r$.
\STATE During the reductions, we use Algorithm~\ref{alg:catalog} to add atoms and their quotients
into the catalog, and to replace them with colored edges.
\medskip
\STATE Using (P2), we compute all semiregular subgroups $\Gamma_r$ of $\Aut(G_r)$.
\FOR{each semiregular subgroup $\Gamma_r$}
	\STATE Compute the quotient $H_r = G_r / \Gamma_r$.
	\STATE Choose, say, the central block/articulation of $H_r$ as the core.
	\STATE Compute the reduction series $H_r,\dots,H_s$ with respect to the core.
	\STATE During the reductions, we use Algorithm~\ref{alg:catalog} to add atoms into the catalog,
	and to replace them with colored edges.
	\medskip
	\FOR{each guessed position of the core in $H$}
		\STATE Compute the reduction series with lists $H = \calR_0,\dots,\calR_t$.
		\renewcommand{\algorithmicfor}{\textbf{to}}
		\FOR{compute $\calR_{i+1}$ from $\calR_i$}
			\renewcommand{\algorithmicfor}{\textbf{for}}
			\FOR{each proper atom or dipole $A$ in $\calR_i$}
				\IF{$A$ is a proper atom}
					\STATE Replace its pendant elements with the unique pendant edges from their
					lists. If some list contains no pendant edge, halt and test for other choices of
					the core in $H$.
				\ENDIF
				\STATE Replace $A$ with a colored edge using the catalog. Halt and test for other
				choices of the core in $H$ if $A$ is not in the catalog.
			\ENDFOR
			\FOR{each block atom $A$ in $\calR_i$}
				\STATE Replace it by a pendant element $x$ whose list $\frakL(x)$ is computed using
				Algorithm~\ref{alg:computing_list}. Halt and test for other choices of the core in
				$H$ if $\frakL(x)$ is empty.
			\ENDFOR
		\ENDFOR
		\medskip
		\STATE Test $\calR_t \listiso H_s$ using Lemma~\ref{lem:testing_expandibility}.
		\IF{$\calR_t \listiso H_s$}
			\STATE Using the lists, compute the expansions $H_{s-1},\dots,H_0$ such that $H_0 \cong H$.
			\STATE Using Theorem~\ref{thm:quotient_expansion}, compute the group extensions
			$\Gamma_{r-1},\dots,\Gamma_0 = \Gamma$ to the interiors of expanded edges, loops and half-edges.
			For half-edges, use involutions $\tau$ on interiors of replacing half-quotients.
			\STATE By Lemma~\ref{lem:expandability_correctness}, $G / \Gamma \cong H$, so the group
			$\Gamma$ defines the regular covering projection $p : G \to H$.
			\medskip
			\RETURN The semiregular subgroup $\Gamma \le \Aut(G)$.
		\ENDIF
	\ENDFOR
\ENDFOR
\medskip
\RETURN The graph $G$ does not regularly cover the graph $H$.
\end{algorithmic}
\end{algorithm}

Next, we prove two corollaries. The first corollary states that if $G$ is $3$-connected, $H$ is
$2$-connected or $k = |G|/|H|$ is odd, the meta-algorithm can avoid the slow subroutine of
Lemma~\ref{lem:lists_for_star_atoms} and can be modified to run in polynomial time.

\begin{proof}[Corollary~\ref{cor:simple_cases}]
If $G$ is 3-connected, then it is primitive, so $G = G_r$. Therefore, we compute all quotients $H_r
= G_r / \Gamma_r$, and test using Lemma~\ref{lem:testing_gi} whether $H_r \cong H$. No reduction
with lists needs to be applied. If $H$ is 2-connected, no pendant elements are created the reduction
of $H$ with lists, so the slow subroutine can be avoided and even the assumptions (P1), (P2) and
(P3${}^*$) are sufficient.

If $|\Gamma| = |\Gamma_r|$ is odd, then no half-edges occur in $H_r$, and so according to
Corollary~\ref{cor:unique_expansion}, the expansion gives the unique graph $H_0$. We just test
whether $H_0 \cong H$.
\end{proof}

Next, we prove that we can modify the meta-algorithm to output all regular quotients of $G$, with a
polynomial-time delay.

\begin{proof}[Corollary~\ref{cor:listing_all_quotients}]
We compute the reduction series $G = G_0,\dots,G_r$ and all semiregular subgroups $\Gamma_r$ of
$\Aut(G_r)$. Next, we run all possible expansions of $H_r = G_r / \Gamma_r$ to $H_0$ using
Theorem~\ref{thm:quotient_expansion}, by all possible choices of half-quotients. All half-quotients
of proper atoms can be computed in polynomial time. For dipoles, we can easily generate them with
polynomial-time delays. We output all constructed graphs $H_0$. 
\end{proof}

\section{Star Blocks Atoms with Lists} \label{sec:star_atoms}

The bottleneck in the running time of the meta-algorithm of Theorem~\ref{thm:metaalgorithm} is the
single slow subroutine in Lemma~\ref{lem:lists_for_star_atoms}, computing lists of pendant elements
replacing star block atoms in $\calR_i$. In this section, we give insights into this problem,
which might lead to a faster algorithm for \rcover of planar graphs.

We show a combinatorial reformulation to finding a certain generalization of a perfect matching
which we call \ivmatch.  Here we describe a complete derivation of this problem, and in Conclusions
we just give its combinatorial statement.

\begin{figure}[t!]
\centering
\includegraphics{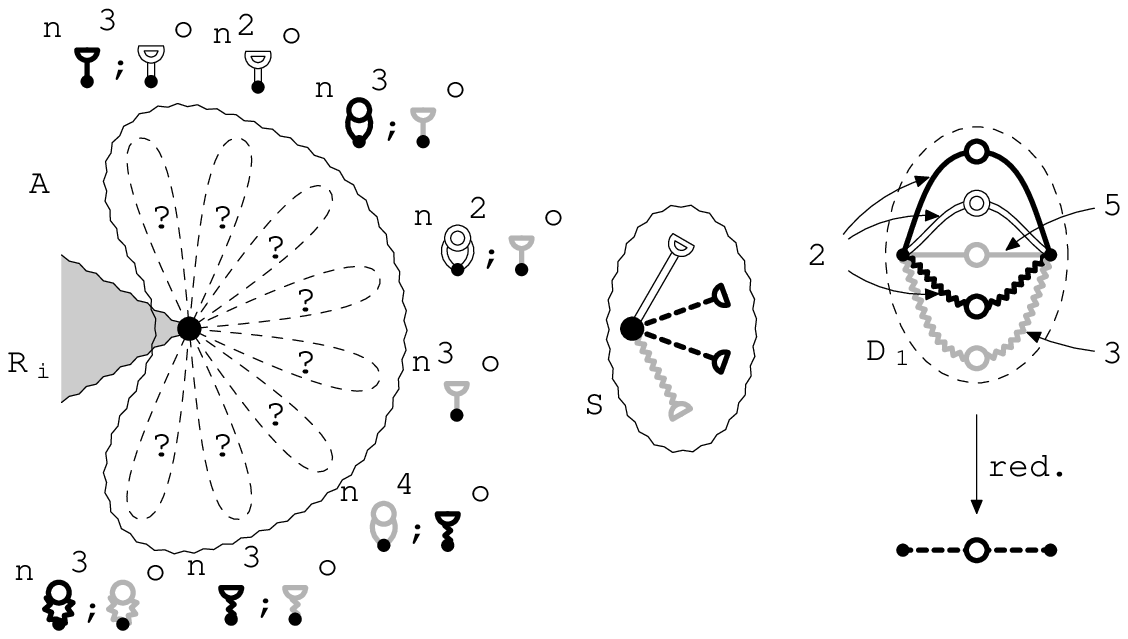}
\caption{On the left, a star block atom $A$ in $\calR_i$ with 23 attached pendant elements, together
with their lists and multiplicities. On the right, a star block atom $S$ from the catalog which
belongs to $\frakL(x)$. The bold dashed edges correspond to the partially expanded dipole $D_1$
whose edges are depicted with multiplicities, the remaining colored edges correspond to proper
atoms.}
\label{fig:testing_for_star_atoms}
\end{figure}

\begin{figure}[b!]
\centering
\includegraphics{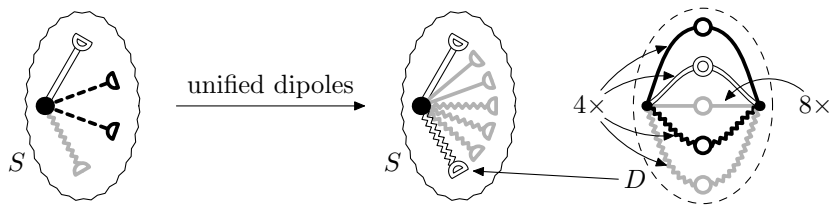}
\caption{We apply the unification on the example in Fig.~\ref{fig:testing_for_star_atoms} as
follows. We unify both occurances of the dipole $D_1$ into the dipole $D$. Since the colored classes
of gray edges have odd sizes, we attach one half-edge per dipole from each class directly to $S$.}
\label{fig:result_of_preprocessing}
\end{figure}

\heading{Instance.}
Figure~\ref{fig:testing_for_star_atoms} shows an example. Suppose that $\calR_i$ contains a star
block atom $A$ with attached pendant elements, each with a previously computed list and
corresponding to a non-star block atom.  We want to determine the list $\frakL(x)$ of the pendant
element $x$ replacing $A$ in $\calR_{i+1}$. The following are the candidates for members of $\frakL(x)$:
\begin{packed_itemize}
\item Each loop corresponding to the loop-quotient of a dipole $D$.
\item Each half-edge corresponding to a half-quotient of a dipole $D$.
\item Each edge corresponding to the edge-quotient of a star block atom $S$.
\end{packed_itemize}

Since loop-quotients of dipoles are uniquely determined, we can easily test them and we can ignore
them. The case of a half-quotient of a dipole $D$ can be reduced to a star block atom $S$ with a
single half-edge attached corresponding to a half-quotient of $D$. If $S$ can be matched to $A$, we
instead add the half-edge corresponding to a half-quotient of $D$ to $\frakL(x)$.
Therefore, in the remainder of this section, we only deal with the case of a star block atoms $S$.
We want to decide whether the edge corresponding to the edge-quotient of $S$ belongs to $\frakL(x)$.
We shall assume that at least one half-edge attached in $S$ corresponds to a dipole, otherwise the
problem is trivial.

\heading{Outline.} In Section~\ref{sec:preprocessing}, we simplify both $A$ and $S$. In
Section~\ref{sec:sizes}, we further apply size constraints to simplify them. In
Section~\ref{sec:iv_match}, we derive the \ivmatch problem.

\subsection{Preprocessing Star Block Atoms} \label{sec:preprocessing}

The star block atom $S$ has several pendant edges, loops and half-edges attached.  Further, we may
assume that $S$ is partially expanded (see Section~\ref{sec:catalog}), so its pendant edges
correspond to non-star block atoms and its loops correspond to proper atoms.  On the other hand, a
half-edge can be of two types: either it corresponds to a half-quotient of a proper atom, or of a
dipole.  For example, in Fig.~\ref{fig:testing_for_star_atoms} we have two half-edges
corresponding to proper atoms, and two half-edges corresponding to dipoles. Further, we may assume
that all these dipoles are partially expanded (see Section~\ref{sec:catalog}), so all their edges
correspond to proper atoms.

\begin{figure}[b!]
\centering
\includegraphics{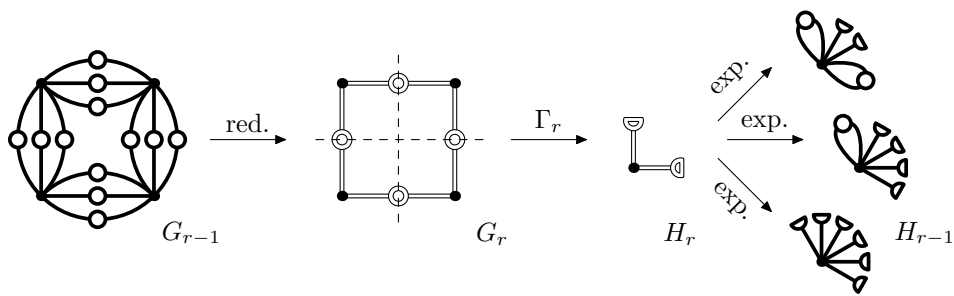}
\caption{For $\Gamma_r$ generated by two reflections, the quotient $H_r$ consists of a star block atom
with two half-edges corresponding to dipoles. All three expansions $H_{r-1}$ up to isomorphism are
depicted. But it is not possible to expand $H_r$ to the quotient with three attached loops.}
\label{fig:star_atom_with_two_dipoles}
\end{figure}

\heading{Unifying Dipoles.}
Recall that every half-quotient of a dipole consists of half-edges and loops attached to one vertex.
Since $S$ may contain multiple half-edges corresponding to half-quotients of dipoles, we want to
unify them into one dipole $D$ containing all their edges. (This may happen in the quotients; see
Fig.~\ref{fig:star_atom_with_two_dipoles}.) 

The issue is that this unification might introduce additional quotients of $D$ as in
Fig.~\ref{fig:star_atom_with_two_dipoles}. If two dipoles both contain an odd number of edges of
one color, the unified dipole $D$ has a half-quotient consisting of only loops of this color which
is not possible in the case of half-quotients of two separated dipoles.  There is an easy fix: we
check each dipole and we remove one edge from each color class of odd size (of necessarily halvable
edges) and attach the half-edge of this color directly to $S$. At least one half-edge of this color
appears in every half-quotient of this dipole, so the possible half-quotients are not changed. In
Fig.~\ref{fig:result_of_preprocessing}, we illustrate this preprocessing for the example in
Fig.~\ref{fig:testing_for_star_atoms}.

\heading{Non-halvable Edges of The Dipole.}
If the dipole $D$ contains some non-halvable edges, then they are paired in every half-quotient of
$D$ and form loops. We remove them from $D$ and attach the corresponding number of loops
directly in $S$. After this step, the dipole $D$ contains only even number of halvable edges in each
color class.

\heading{Attached Pendant Edges and Loops.}
The star block atom $S$ may have some pendant edges (corresponding to non-star block atoms) and loops
(corresponding to proper atoms) attached. Therefore, each attached pendant edge/loop corresponds to
exactly one pendant element of $A$. By Lemma~\ref{lem:list_properties}, each is contained in list of
only one type of pendant elements, all corresponding to isomorphic block parts in $H$. Therefore,
we can arbitrarily assign pendant elements, remove them from $A$ and remove these pendant edges and
loops from $S$. 

\heading{Summary.}
By the preprocessing of $S$ and $A$ described above, we may assume the following.  The star
block atom $S$ has only half-edges attached, all but one corresponding to proper atoms. The
remaining half-edge corresponds to the unified dipole $D$ having only color classes of even sizes of
halvable edges corresponding to proper atoms.

For each pendant element $x$ of $A$, the list $\frakL(x)$ contains only half-edges (attached in $S$
or in a half-quotient of $D$) and loops (corresponding to halvable edges of $D$).

\subsection{Sizes and Chains} \label{sec:sizes}

To simplify the problem further, we study sizes of atoms and their quotients.  Let $A$ be an atom
and let $Q$ be a quotient of this atom. Depending on the type of $Q$, we get:
\begin{packed_itemize}
\item \emph{$Q$ is the edge-quotient:} Then $\bv(Q) = \bv(A)$ and $\be(Q) = \be(A)$.
\item \emph{$Q$ is the loop-quotient:} Then $\bv(Q) = \bv(A)-1$ and $\be(Q) = \be(A)$.
\item \emph{$Q$ is a half-quotient:} Then $\bv(Q) = \bv(A)/2$ and $\be(Q) = \be(A)/2$.
\end{packed_itemize}

\heading{Sizes of Expanded Subgraphs and Quotients.}
Throughout each reduction, we calculate how many vertices and edges are in all the atoms replaced by
colored edges which we denote by $\bhatv$ and $\bhate$. Initially, we put $\bhatv(e) = 0$ and
$\bhate(e) =
1$ for every edge $e \in \bE(G_0)$. For a subgraph $X$, we define
$$
\bhatv(X) := \bv(X) +\!\!\sum_{e \in \bE(X)}\!\! \bhatv(e),
\qquad\text{and}\qquad
\bhate(X) := \!\! \sum_{e \in \bE(X)}\!\! \bhate(e).
$$
When an atom $A$ is replaced by an edge $e$ in the reduction, we put $\bhatv(e) = \bhatv(\int A)$
and $\bhate(e) = \bhate(\int A)$. For a subgraph $X$ of $G_i$, the numbers $\bhatv(X)$ and
$\bhate(X)$ are the numbers of vertices and edges when $X$ is fully expanded.

We similarly define $\bhatv$ and $\bhate$ for quotients and their subgraphs; the difference is that
the quotients may contain half-edges. For a half-edge $h \in \bH(X)$, created by halving an edge $e$,
we put $\bhatv(h) = \bhatv(e)/2$ and $\bhate(h) = \bhate(e)/2$. For a subgraph $X$, we define
$$
\bhatv(X) := \bv(X) + \!\!\sum_{e \in \bE(X)}\!\! \bhatv(e) + \!\!\sum_{h \in \bH(X)}\!\! \bhatv(h),
\qquad\text{and}\qquad
\bhate(X) := \!\!\sum_{e \in \bE(X)}\!\! \bhate(e) + \!\!\sum_{h \in \bH(X)}\!\! \bhate(h).
$$

\heading{Sizes of Pendant Elements.}
We also inductively define $\bhatv$ and $\bhate$ for pendant elements $x$ and subgraphs $X$ of
$\calR_0,\dots,\calR_t$. Initially, we put $\bhatv(e) = 0$ and $\bhate(e) = 1$ for every edge $e \in
\bE(\calR_0)$. For a subgraph $X$ of $\calR_i$, let $\bP(X)$ be the set of all pendant elements in $X$.
We define 
$$
\bhatv(X) := \bv(X) + \!\!\sum_{e \in \bE(X)}\!\! \bhatv(e) + \!\!\sum_{y \in \bP(X)}\!\! \bhatv(\int y),
\qquad\text{and}\qquad
\bhate(X) := \!\!\sum_{e \in \bE(X)}\!\! \bhate(e) + \!\!\sum_{y \in \bP(X)}\!\! \bhate(y),
$$
while for a pendant element $x$ corresponding to an atom $A$ in $\calR_i$, we define $\bhatv(x) =
\bhatv(A)$, $\bhatv(\int x) = \bhatv(\int A) = \bhatv(x)-1$, and $\bhate(x) = \bhate(A)$.

\heading{Restricting Lists by Sizes.} Next, we show that these sizes can restrict possible members
of lists of pendant elements:

\begin{lemma} \label{lem:list_sizes}
For a pendant element $x$, the possible pendant edge and all loops and half-edges of the list
$\frakL(x)$ have the same $\bhatv$ and $\bhate$ as $\bhatv(x)$ and $\bhate(x)$, respectively.
\end{lemma}

\begin{proof}
The pendant element $x$ corresponds to a block part of $H$.  All members of $\frakL(x)$ can be fully
expanded to graphs isomorphic to this block part.  Necessarily, these graphs contain the same number
of vertices and edges as $\bhatv(x)$ and $\bhate(x)$. 
\end{proof}

When $\frakL(x)$ is computed, we can only consider quotients of the correct sizes, speeding up
Algorithm~\ref{alg:computing_list}. For pendant edges and loops, each belongs to lists of only one
type of pendant elements by Lemma~\ref{lem:list_properties}. This is not true for half-edges and 
for purpose of this section, the following is important:

\begin{corollary}
Let $x$ and $y$ be two pendant elements.
\begin{packed_enum}
\item[(i)] If $\frakL(x)$ and $\frakL(y)$ share a half-edge, then $\bhatv(x) = \bhatv(y)$ and
$\bhate(x) = \bhate(y)$.
\item[(ii)] Let $\frakL(x)$ contain a loop of a color $c$ and a half-edge of a color $c'$. Then
$\frakL(y)$ cannot contain both the loop of the color $c'$ and the half-edge of the color $c$.
\end{packed_enum}
\end{corollary}

\begin{proof}
(i) Implied by Lemma~\ref{lem:list_sizes}.

(ii) Let $\frakL(x)$ contain a loop $e$ and $\frakL(y)$ contain a half-edge $h$ of the same color.
Then $\bhatv(x) = \bhatv(e)+1$ and $\bhatv(y) = \bhatv(e)/2+1$ for the vertices, and $\bhate(x) =
\bhate(e)$ and $\bhate(y) = \bhate(e)/2$ for the edges. Therefore $x$ corresponds to a larger block
part in $H$ than $y$. By the same argument, we deduce that $y$ corresponds to a larger block part in
$H$ than $x$, which gives a contradiction.
\end{proof}

The property (i) relates half-edges together.  The property (ii) states that there is a certain size
hierarchy on the pendant elements discussed below.

\begin{figure}[t!]
\centering
\includegraphics{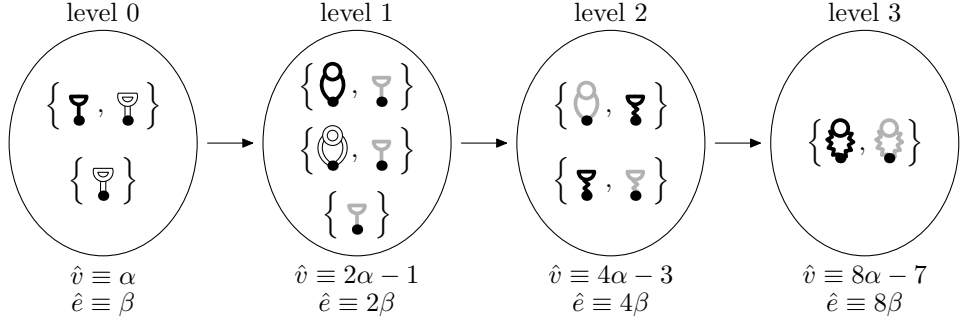}
\caption{A chain of pendant elements with four levels, which is the only chain in
Fig.~\ref{fig:testing_for_star_atoms}, for some $\alpha$ and $\beta$ (we ignore multiplicities of
pendant elements). Notice that quotients corresponding to one atom are placed in neighboring
levels.}
\label{fig:chains_of_pendant_elements}
\end{figure}

\heading{Chains of Pendant Elements.}
Pendant elements can be partitioned into independent chains, each further partitioned into several
levels. The level of size $(\alpha,\beta)$ consists of all pendant elements $x$ having $\bhatv(x) =
\alpha$ and $\bhate(x) = \beta$. Each chain starts with the level $0$ of some size
$(\alpha,\beta)$. Further, it contains the levels $m>0$ of sizes $(2^m\alpha - (2^m-1),2^m \beta)$.
See Fig.~\ref{fig:chains_of_pendant_elements} for an example.

The key property is the following: if $\frakL(x)$ contains a half-edge of a color $c$ and
$\frakL(y)$ contains the loop of the same color $c$, then $x$ belongs to a level $m$ and $y$ belongs
to the level $m+1$ of the same chain. A star block atom $A$ can contain multiple chains, but
different chains contain completely different colors in their lists, so they are completely
independent. 

\heading{Summary.} We may partition $S$, $D$, and $A$ according chains of pendant elements of $A$ and test
them separately. Therefore, we may assume that there is exactly one chain of pendant elements in
$A$, and only the corresponding edges in $D$ and half-edges in $S$.

\subsection{Reduction to the IV-Matching Problem} \label{sec:iv_match}

The star block atom $S$ contains a half-edge corresponding to the dipole $D$ and let $\bH'(S)$ be the
set of the remaining half-edges corresponding to proper atoms. The dipole $D$ has the following
half-quotients $Q$. For each color class of an even size $s$, we choose an arbitrary integer $\ell$
such that $0 \le \ell \le {s \over 2}$, and attach $\ell$ loops and $s-2\ell$ half-edges of this
color in $Q$.

If these values $s$ and $\ell$ are known for each color class, we can test existence of a perfect
matching as described in the proof of Lemma~\ref{lem:lists_for_star_atoms}. Since they are not
known, we need to solve a generalization of perfect matching called \ivmatch
which includes choosing of these values as a part of the problem.

\heading{Definition of the Problem.}
The input of \ivmatch\ gives the following bipartite graph $B$. We have $\bV(B) = \bP(A) \cup \bE(D) \cup
\bH'(S)$.  For $e \in \bE(D)$ of a color $c$ and $x \in \bP(A)$, we have $ex \in \bE(B)$ if and only if the
half-edge or the loop of the color $c$ belongs to the list $\frakL(x)$. We call the former case a
\emph{half-incidence} and the latter case a \emph{loop-incidence}. Further, we have a
\emph{half-incidence} $hx \in \bE(B)$ between $h \in \bH'(S)$ of a color $c$ and $x \in \bP(A)$ if and
only if the half-edge of the color $c$ belongs to $\frakL(x)$. 

\begin{figure}[t!]
\centering
\includegraphics{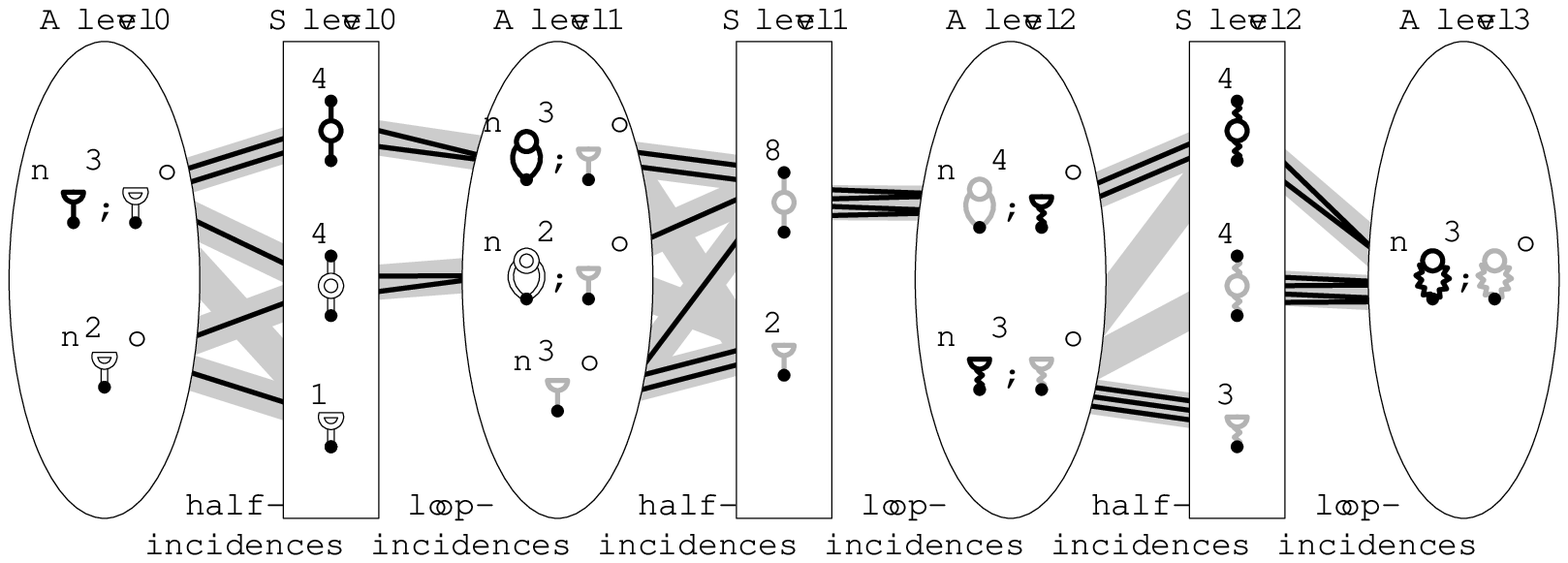}
\caption{The instance of the \ivmatch\ problem corresponding to the input $A$ in
Fig.~\ref{fig:testing_for_star_atoms} and the preprocessed star block atom $S$ in
Fig.~\ref{fig:result_of_preprocessing}. The edges $\bE(B)$ are depicted in gray and an \ivsubgraph
$B'$ is highlighted in bold. We have {\sffamily I}'s between half-incidences and {\sffamily V}'s
between loop-incidences.  The part $\bP(A)$ is in elipses, the other part $\bE(D) \cup \bH'(S)$ is
in boxes. The lists $\bP(A)$, the edges $\bE(D)$ and the half-edges $\bH'(S)$ are depicted together with
multiplicities.}
\label{fig:v-matching_problem_example}
\end{figure}

We ask whether there exists a \emph{spanning subgraph} $B'$ of $B$, called an \ivsubgraph of $B$,
satisfying the following properties. Each component of connectivity of $B'$ is a path of length one
or two (corresponding to \textsf{I} and \textsf{V} in the name). Each $x \in \bP(A)$ is in $B'$ either
half-incident to exactly one vertex in $\bE(D) \cup \bH'(S)$, or it is loop-incident to exactly two
edges $e,e' \in \bE(D)$ of the same color class.  Further, each vertex of $\bE(D) \cup \bH'(S)$ is
incident in $B'$ to exactly one $x \in \bP(A)$. See Fig.~\ref{fig:v-matching_problem_example} for an
example, with several additional properties which we discuss below.

\heading{Level Structure.}
The structure of levels of the chain of pendant elements transfers into the level structure of
$B$. The part $\bP(A)$ is partioned into levels called \emph{$A$ levels}. Every half-edge $h \in \bH'(S)$ is
half-incident only to vertices of the $A$ level $m$ of the size $(\bhatv(h),\bhate(h))$. Every edge
$e \in \bE(D)$ is half-incident only to vertices of the $A$ level $m$ of size
$(\bhatv(e)/2,\bhate(e)/2)$ and loop-incident only to vertices of the $A$ level $m+1$ of the size
$(\bhatv(e)-1,\bhate(e))$.  Therefore, we can define \emph{$S$ levels} for the part $\bE(D) \cup
\bH'(S)$ such that an $S$ level $m$ contains all vertices half-incident to the $A$ level $m$ or
loop-incident to the $A$ level $m+1$.  If we depict all levels from left to right according to their
order, alternating $A$ levels and $S$ levels, all edges of $B$ go between consecutive levels as
depicted in Fig.~\ref{fig:v-matching_problem_example}.

\heading{Clusters.}
We can view $B$ as a cluster graph. In each $S$ level, $\bE(D)$ and $\bH'(S)$ form clusters according to
their color classes, called \emph{edge clusters} and \emph{half-edge clusters} respectively. In each
$A$ level, the pendant elements form \emph{pendant element clusters} according to equivalence
classes of their lists. (We note that two pendant elements with equal lists can correspond to
non-isomorphic subgraphs in $H$. Then their lists contain only half-edges.) Two clusters are either
completely adjacent, or not adjacent at all: the subgraph induced by the union of two clusters
is either a complete bipartite graph, or contains no edges.

Each pendant element cluster may be loop-adjacent to several edge clusters. On the other hand, each
edge cluster is loop-adjacent to at most one pendant element cluster: the loop of the
corresponding color belongs to at most one isomorphism class of pendant elements by
Lemma~\ref{lem:list_properties}.  There are no constraints for half-incidences between clusters.

\heading{Only Logarithmically Many Levels.}
The sizes of graphs are growing exponentially with the level number. Therefore, we have at most
logarithmically many levels with respect to the size of the input graph $H$.

\heading{Complexity.}
Since \ivmatch can be used to solve the slow subroutine of Lemma~\ref{lem:lists_for_star_atoms},
we get the following in relation to the meta-algorithm of Theorem~\ref{thm:metaalgorithm}:

\begin{proposition} \label{prop:ivmatch}
If the \ivmatch problem can be solved for logarithmically many levels in polynomial time, then we can
modify the meta-algorithm of Theorem~\ref{thm:metaalgorithm} to run in polynomial time as well.
\end{proposition}

Unfortunately, Folwarczn\'y and Knop~\cite{iv_matching} recently proved that this problem is
\cNP-complete, even for two $A$ levels. Nevertheless, we believe that the particular instances
arizing from the \rcover problem might be solvable in polynomial time and the properties
described in this section might be useful for that.

\heading{Numbers of Edges Between Levels.}
We do not know how many half- and loop-incidences are in $B'$ at each cluster, otherwise we could
solve the problem directly by finding a perfect matching in a modified graph.  On the other hand,
these numbers are determined between consecutive $A$ levels and $S$ levels as follows.  Let $a_m$ be
the number of pendant elements in the $A$ level $m$ and let $s_m$ be the number of edges and
half-edges in the $S$ level $m$. Let $B'$ be an \ivsubgraph and let $b_i$ and $b_i'$ be the numbers
of edges in $B'$ between the $A$ level $i$ and the $S$ level $i$, and between the $S$ level $i$ and
the $A$ level $i+1$, respectively. 

Every pendant element in the $A$ level $0$ is half-incident in $B'$ to the
$S$ level $0$, so $b_0 = a_0$. Therefore, the number of remaining vertices in the $S$ level $0$ is
$s_0 - b_0$. These vertices are loop-incident in $B'$ to the $A$ level $1$, so $b'_0 = s_0 - b_0$
and $a_1 - {b'_0 \over 2}$ pendant elements remain in the $A$ level $1$. We can proceed in this way
further, and we get the following inductive formulas:
$$b_i = a_i - {b'_{i-1} \over 2},\qquad\text{and}\qquad b'_i = s_i - b_i.$$
Clearly, each number $b'_i$ has to be even, otherwise no \ivsubgraph exists. 

\section{Applying the Meta-algorithm to Planar Graphs} \label{sec:planar_graphs}

In this section, we discuss automorphism groups of 3-connected planar graphs and we show that the
meta-algorithm of Theorem~\ref{thm:metaalgorithm} applies to the class of planar graphs.

\begin{figure}[t!]
\centering
\includegraphics{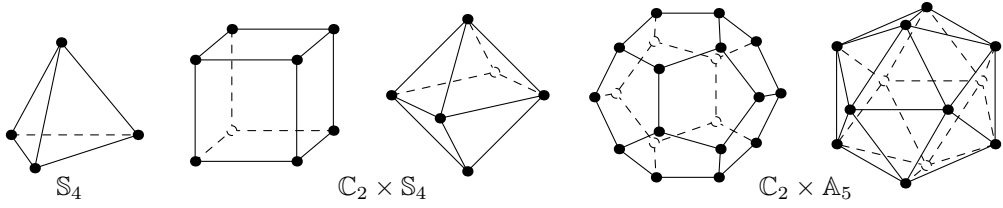}
\caption{The five platonic solids together with their automorphism groups.}
\label{fig:platonic_solids}
\end{figure}

\heading{Automorphism Groups.}
A group is \emph{spherical} if it is the group of the symmetries of a tiling of the sphere. The
first class of spherical groups are the subgroups of the automorphism groups of the platonic solids,
i.e., $\gS_4$ for the tetrahedron, $\gC_2 \times \gS_4$ for the cube and the octahedron, and $\gC_2
\times \gA_5$ for the dodecahedron and the icosahedron; see Fig.~\ref{fig:platonic_solids}.  The
second class of spherical groups is formed by the infinite families $\gC_n$, $\gD_n$, $\gC_n \times
\gC_2$, and $\gD_n \times \gC_2$.

A map $\calM$ is a 2-cell embedding of a graph $G$ onto the sphere.  A \emph{rotation} at a vertex
is a cyclic ordering of the edges incident with the vertex. An \emph{angle} is a triple $(v,e,e')$
where $v$ is a vertex, and $e$ and $e'$ are two incident edges which are consecutive in the rotation
at $v$ or in the inverse rotation at $v$. An automorphism of a map is an automorphism of the graph
which preserves the angles; in other words the rotations are preserved.  For a 3-connected planar
graph $G$, we have $\Aut(G) \cong \Aut(\calM)$ and it is a spherical group.  For more details,
see~\cite{fkkn16} and the references therein.

\begin{lemma} \label{lem:3conn_planar_aut}
For a $3$-connected planar graph $G$, we can compute $\Aut(G)$ in time $\O(\bv^2(G))$.
\end{lemma}

\begin{proof}
Compute the unique map $\calM$ of $G$. There are $\O(\bv(G))$ angles in $\calM$. We fix one angle
$(v,e,e')$, and test for each other angle whether there is an automorphism mapping $(v,e,e')$ to it.
The key observation is that this partial mapping has a unique extension to a mapping compatible with
the rotations of $\calM$. We can just test in $\O(\bv(G))$ whether it is an automorphism.  The total
running time is $\O(\bv^2(G))$.
\end{proof}

\heading{Properties (P1) to (P3).} We are ready to establish the following:

\begin{lemma} \label{lem:planar_graph_properties}
The class of planar graphs satisfies (P1) to (P3).
\end{lemma}

\begin{proof}
The class of planar graphs clearly satisfies (P1). For (P2), Lemma~\ref{lem:3conn_planar_aut} allows
to compute $\Aut(G)$ in time $\O(\bv^2(G))$. Since it is a spherical group, we can generate all
linearly many subgroups and check which ones act semiregularly.  The property (P3) holds for
projectively planar graphs since \lgi can be solved in time $\O(\bv^{5/2}(G))$ for graphs of bounded
genus~\cite{kkz} (even for lists on both vertices and half-edges).
\end{proof}

\begin{proof}[Theorem~\ref{thm:planar_rcover}]
By Lemma~\ref{lem:planar_graph_properties}, we can apply Theorem~\ref{thm:metaalgorithm}.
\end{proof}

\begin{figure}[t!]
\centering
\includegraphics[scale=0.95]{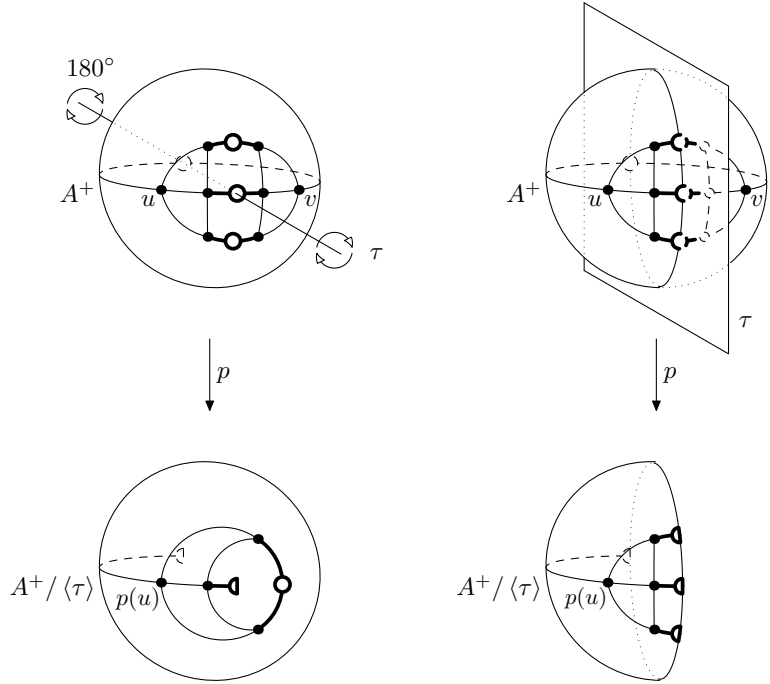}
\caption{The rotational quotient and reflectional quotient of a planar proper atom $A$ with the added edge
$uv$.}
\label{fig:halfquotients_of_proper_atoms}
\end{figure}

\heading{Half-quotients of Planar Proper Atoms.}
Let $A$ be a planar proper atom. We know that $\Aut(A)$ is a spherical group, and 
further each semiregular involution $\tau$ defining a half-quotient of $A$ has to 
exchange the vertices of the boundary. We get two types of geometrically defined quotients, depicted
in Fig.~\ref{fig:halfquotients_of_proper_atoms}.

\begin{lemma}[\cite{fkkn16}, Lemma 5.6] \label{lem:planar_proper_half_quotients}
Let $A$ be a planar proper atom and let $\bo A = \{u,v\}$. There are at most two half-quotients $A /
\left<\tau\right>$ where $\tau \in \Aut(A)$ is an involutory semiregular automorphism
transposing $u$ and $v$:
\begin{packed_enum}
\item[(a)] \emph{The rotational half-quotient} -- The involution $\tau$ is orientation preserving
and $A / \left<\tau\right>$ is planar with at most one half-edge.
\item[(b)] \emph{The reflectional half-quotient} -- The involution $\tau$ is a reflection and $A /
\left<\tau\right>$ is planar with at least two half-edges.
\end{packed_enum}
\end{lemma}

\section{Concluding Remarks} \label{sec:conclusions}

This paper is based on the structural results of~\cite{fkkn16}, describing behaviour of regular
graph covering with respect to 1-cuts and 2-cuts in $G$. In Theorem~\ref{thm:metaalgorithm}, we
derive an FPT meta-algorithm for testing regular graph covers for $\calC$-inputs $G$ where $\calC$
is a class of graphs satisfying (P1) to (P3). In particular, this meta-algorithm is tailored for the
class of planar graphs (Theorem~\ref{thm:planar_rcover}).

When working with 3-connected decomposition, we described two subroutines we need to solve. First,
we have rediscovered graph isomorphism restricted by lists, introduced by Lubiw~\cite{lubiw}, which
lead to several fruitful results in~\cite{kkz}. Second, we introduce a generalization of bipartite
matching called the \ivmatch problem, proved to be \cNP-complete by Folwarcn\'y and
Knop~\cite{iv_matching}.

We conclude by several remarks and open problems.

\heading{Running Time of The Meta-algorithm.}
We have omitted polynomial factors in the complexity since the main goal was to establish that
\rcover can be solved in FTP time for $\calC$-inputs $G$ for $\calC$ satisfying (P1) to (P3).
The degree of the polynomial depends on the complexity of polynomial time algorithms in (P2) and
(P3).

We roughly estimate the degree of the polynomial for running time of the algorithm of
Theorem~\ref{thm:planar_rcover} for planar graphs. Let $n = \bv(G) \ge \bv(H)$. For planar graphs,
each primitive graph has $\O(n)$ quotients and each proper atom has at most two half-quotients.
Therefore, the catalog contains $\O(n^2)$ atoms and quotients. Each catalog query can be answered in
time $\O(n^4)$, and with a suitable canonization even in time $\O(n^3)$.

The reduction series can be computed in time $\O(n)$ by Hopcroft and
Tarjan~\cite{hopcroft_tarjan_dividing}, the symmetry type of each atom can be determined in time
$\O(n^2)$ by Lemma~\ref{lem:3conn_planar_aut}.
When adding a proper atom to the catalog, we compute its at most two half-quotients
in time $\O(n^2)$ and compute their reduction series in time $\O(n^4)$.
Together with catalog queries, we can compute the reduction series and $G_r$ in time $\O(n^5)$.

Next, we iterate over $\O(n)$ quotients $H_r$ of $G_r$. For each, we compute the reduction
series in time $\O(n^4)$. Next, we iterate over $\O(n)$ choices of the core in $H$. We compute the
reduction series with lists where $\O(n)$ subroutines are called. The subroutine of
Lemma~\ref{lem:lists_for_nonstar_atoms} runs in time $\O(n^{9/2})$ since \lgi takes time
$\O(n^{5/2})$~\cite{kkz} and we compare each non-star block atom $A$ with $\O(n^2)$ candidates from
the catalog. The subroutine of Lemma~\ref{lem:lists_for_star_atoms} runs in time $\O(n^2
2^{\be(H)/2})$. The final test in Lemma~\ref{lem:testing_expandibility} runs in time $\O(n^{5/2})$.

In total, the running time of the algorithm in Theorem~\ref{thm:planar_rcover} is
$\O(n^5 \cdot (n^{5/2}+2^{\be(H)/2}))$. By a more careful analysis of the subroutines and possibly
reordering them, it should be possible to decrease the degree little bit.

We did not try to optimize the factor $2^{\be(H)/2}$. This estimate is certainly very rough
and maybe some further techniques from parameterized complexity can be applied to solve \ivmatch
faster. We believe that this approach should be followed only when we can prove that \rcover is
\cNP-complete for planar inputs $G$. Also, to prove Theorem~\ref{thm:metaalgorithm}, it might be
possible to design a simpler FPT algorithm running in time $\O^*(2^{\be(H)/2})$. Our goal was to
obtain as much understanding of the problem as possible, in order to construct a polynomial time
algorithm. We failed with the subroutine of Lemma~\ref{lem:lists_for_star_atoms}, but nevertheless
further structural results may be established and the problem may be solvable in polynomial time.

\heading{Possible Extensions of The Meta-algorithm.}
There are several possible natural extensions of the meta-algorithm. First, we can easily generalize
it for input graph $G$ and $H$ with half-edges, directed edges and halvable edges, and also for
colored graphs. Further, for a regular covering testing, one can prescribe a list $\frakL(u)
\subseteq \bV(H)$ of allowed images of a regular covering projection $p$ for each vertex $u \in
\bV(G)$ such that $p(u) \in \frakL(u)$: the expandability testing subroutine can compute with these
lists as well.

\heading{Complexity of Regular Graph Covering.}
By Lemmas~\ref{lem:rcover_in_np} and~\ref{lem:gi_hardness}, it follows that \rcover is \cGI-hard and
belongs to \cNP. Its complexity remains an open problem:

\begin{problem}
What is the complexity of the \rcover problem?
\end{problem}

\noindent One possibility to attack this problem would be to prove that it is \cNP-hard, or to
construct an efficient algorithm using an oracle for \gi. If \rcover is not \cNP-hard, another
possibility is to prove that \rcover satisfies some properties which unlikely hold for any \cNP-hard
problem.  For instance for the graph isomorphism problem, there are currently three evidences that
it is unlikely \cNP-complete: equivalence of existence and counting~\cite{babai77,mathon_isocount},
\gi belongs to \ccoAM, so polynomial-hierarchy would collapse if \gi is
\cNP-complete~\cite{goldreich1986,schoning1988graph}, and \gi can be solved in subexponential
time~\cite{babai_quasipoly}. 

As a possible next direction of research, we suggest to attack classes of graphs close to planar
graphs, for instance projective planar graphs or toroidal graphs. To do so, it seems that new
techniques need to be built. Even the automorphism groups of projective planar graphs and toroidal
graphs are not yet well understood.

It is natural to ask whether FPT running time of the meta-algorithm of
Theorem~\ref{thm:metaalgorithm} is needed:

\begin{problem}
Can the \rcover problem be solved in polynomial time for $\calC$-inputs $G$ where $\calC$ satisfies
(P1) to (P3)? Can it be solved in polynomial time for planar inputs $G$?
\end{problem}

\heading{The IV-Matching Problem.}
In Section~\ref{sec:star_atoms}, we have shown that the bottleneck of the algorithm of
Theorem~\ref{thm:metaalgorithm} reduces to a generalized matching problem called \ivmatch. Here, we
describe a purely combinatorial formulation of the \ivmatch\ problem. This reformulation can be
useful to understand the problem without regular covering and the structural results obtained
in~\cite{fkkn16} and this paper.

The input of \ivmatch\ consists of a bipartite graph $B$ with a partitioning
$V_1,\dots,V_\ell$ of its vertices $\bV(B)$ which we call \emph{levels}, with all edges
between of consecutive levels $V_i$ and $V_{i+1}$, for $i=1,\dots,\ell-1$. The levels
$V_1,V_3,\dots$ are called \emph{odd} and the levels $V_2,V_4,\dots$ \emph{even}. Further each level
$V_i$ is partitioned into several \emph{clusters}, each consisting of a few vertices with identical
neighborhoods. There are three key properties:
\begin{packed_itemize}
\item The incidences in $B$ respect the clusters; between any two clusters the graph $B$ induces
either a complete bipartite graph, or an edge-less graph.
\item Each cluster of an even level $V_{2t}$ is incident with at most one cluster at $V_{2t+1}$.
\item The incidences between the clusters of $V_{2t-1}$ and $V_{2t}$ can be arbitrary.
\end{packed_itemize}

\begin{figure}[t!]
\centering
\includegraphics{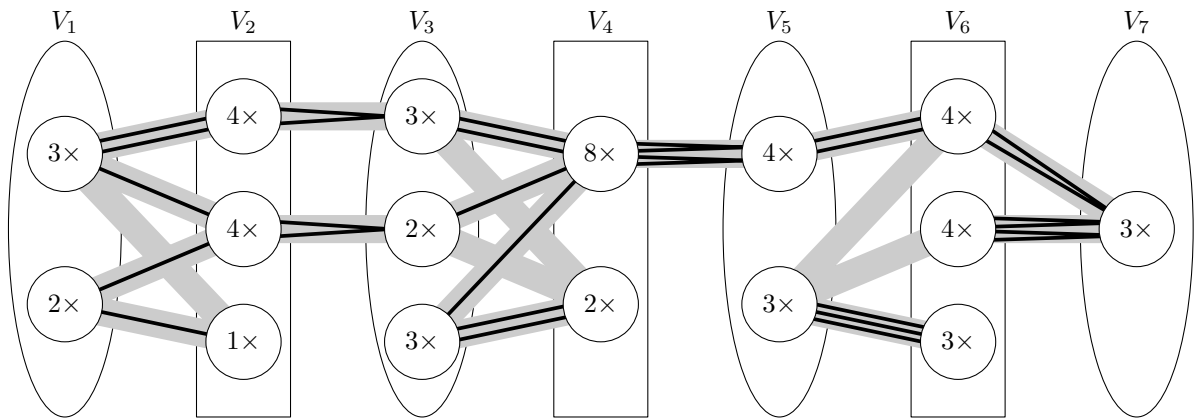}
\caption{An example input $B$, the clusters are depicted by circles together with their sizes.
The odd levels are drawn in circles and the even ones in rectangles. The edges of $B$ are depicted
by gray lines between clusters representing complete bipartite graphs. One spanning subgraph $B'$
solving the \ivmatch\ problem is depicted in bold.}
\label{fig:example_comb_v-matching}
\end{figure}

The problem \ivmatch\ asks whether there is a \emph{spanning subgraph} $B'$ called an \ivsubgraph of
$B$. Each
component of connectivity of $B'$ equals to a path of length one or two. Each vertex of an odd level $V_{2t+1}$ is in $B'$ adjacent either to exactly
one vertex of $V_{2t+2}$, or to exactly two vertices of $V_{2t}$. Each vertex of an even level
$V_{2t}$ is adjacent to exactly one vertex of the levels $V_{2t-1} \cup V_{2t+1}$. In other words,
from $V_{2t-1}$ to $V_{2t}$ the edges of $B'$ form a matching, not necessarily perfect.  From
$V_{2t}$ to $V_{2t+1}$, the edges of $B'$ form independent {\sffamily V}-shapes, with their centers in the
level $V_{2t+1}$. Figure~\ref{fig:example_comb_v-matching} shows an example.

In the conference version of this paper~\cite{fkkn}, we asked as an open problem what is the
complexity of the \ivmatch\ problem.  Recently, Folwarcn\'y and Knop~\cite{iv_matching} answered
this by proving that \ivmatch is strongly \cNP-hard even for $\ell=3$. We note that this does not
imply \cNP-hardness of \rcover, and it is possible that specific instances of \ivmatch arising from
\rcover can be solved in polynomial time.

\heading{A Weaker Assumption (P2').}
To make Theorem~\ref{thm:metaalgorithm} a more natural generalization of Babai's
algorithm~\cite{babai1975automorphism} for graph isomorphism, it would be nice to replace (P2) with
a weaker assumption:
\begin{packed_itemize}
\item[(P2')] For a 3-connected graph $G \in \calC$ with colored vertices and colored possibly
directed edges and a graph $H$ with colored vertices and colored possibly directed edges, half-edges
and loops, we can test $\rcover(G,H)$ in polynomial time.
\end{packed_itemize}
It is an open problem whether our meta-algorithm can be modified for $\calC$ satisfying (P1), (P2')
and (P3):\footnote{Even more natural would be to replace (P3) with (P$3^*$), but this problem seems
even more tricky.}

\begin{problem}
Can the \rcover problem be solved in FPT time (with respect to the parameter $\be(H)$) for
$\calC$-inputs $G$ where $\calC$ satisfies (P1), (P2'), and (P3)?
\end{problem}

The modified algorithm would have to process both $G$ and $H$ simultaneously. For instance, it is
not needed to decide whether a proper atom is halvable or symmetric. Also, we do not need to compute
all half-quotients of a proper atom, we only need to test whether some subgraphs located in $H$ are
one of them. There are several issues with this approach which make this generalization a very
tricky problem.

As illustrated in Fig.~\ref{fig:catalog_quotient_reduction}, a half-quotient of a
proper atom might not be essentially 3-connected (but it is always essentially 2-connected if the
proper atom is not a path). Therefore, we would locate $Q'$ in $H$ which is not a half-quotient of
$A$. To test it using (P2), we would need to expand some proper atoms and dipoles in $Q'$ to reach
$Q$, but it is not clear which ones.

Even more involved is to avoid finding of all quotients $H_r = G_r / \Gamma_r$. Since $H_r$ might
consist of many blocks which might contain many proper atoms and dipoles, it is not clear how to
locate it in $H$ to test existence of a regular covering using (P2).

\bibliographystyle{siam}
\bibliography{algorithmic_regular_covers_II}

\end{document}